\documentclass{emulateapj}
\slugcomment{{\sc Accepted to ApJ:} October 1, 2009}

\shorttitle{FUV Dust Albedo in Upper Sco}
\shortauthors{Lewis et al.}

\newcommand{\shortband}{912-1029 \AA}
\newcommand{\midband}{1030-1200 \AA}
\newcommand{\longband}{1235-1450 \AA}
\newcommand{\hone}{\mbox{\ion{H}{1}}}

\begin{document}

\title{Far-Ultraviolet Dust Albedo Measurements in the Upper 
Scorpius Cloud Using the SPINR Sounding Rocket Experiment}

\author{N. K. Lewis, T. A. Cook, K. P. Wilton, S. Chakrabarti}
\affil{Center for Space Physics and Department of Astronomy, Boston University, \\Boston, MA 02215}
\email{nkhoward@bump.bu.edu}

\author{K. France}
\affil{Canadian Institute for Theoretical Astrophysics, University of Toronto, \\Toronto ON M5S 3H8}
\email{france@cita.utoronto.ca}

\and

\author{K. D. Gordon}
\affil{Steward Observatory, University of Arizona, Tucson, AZ 85721}  
\email{kgordon@as.arizona.edu}

\begin{abstract}
The Spectrograph for Photometric Imaging with Numeric Reconstruction (SPINR) sounding rocket 
experiment was launched on 2000 August 4 to record far-ultraviolet (912-1450 \AA) spectral and spatial information 
for the giant reflection nebula in the Upper Scorpius region.  The data were divided into three arbitrary bandpasses 
(\shortband, \midband, and \longband) for which stellar and nebular flux levels were derived.  These flux 
measurements were used to constrain a radiative transfer model and to determine the dust albedo for the 
Upper Scorpius region.  The resulting albedos were $0.28\pm0.07$ for the \shortband\ bandpass, 
$0.33\pm0.07$ for the \midband\ bandpass, and $0.77\pm0.13$ for the \longband\ bandpass.
\end{abstract}

\keywords{dust, extinction --- ISM: individual (Scorpius OB2) --- radiative transfer 
--- ultraviolet: ISM}

\section{Introduction}

Dust in the interstellar medium (ISM) is central to many physical 
processes, such as star and planet formation.  In the 
ultraviolet (UV), the properties of interstellar dust are largely gleaned 
from the interactions between dust and the electromagnetic radiation from 
UV bright stars.  Typically, these interactions manifest themselves through 
the absorption and scattering of UV light.  The reduction of starlight by 
dust is commonly referred to as extinction, which has been characterized 
in the Milky Way Galaxy by an idealized set of wavelength dependent 
`extinction curves' (see the review by \citet{fit04}).  

The light scattering properties of dust are characterized by two 
quantities:  the albedo ($a$) and the asymmetry parameter ($g$). 
Both $a$ and $g$ are functions of the wavelength 
of the incident light.  Previous studies have 
probed reflection nebulae, dark clouds, and diffuse galactic light 
in an attempt to characterize the scattering properties of interstellar 
dust from infrared (IR) to UV wavelengths (see the review 
by \citet{gor04}).  Although these studies have contributed to 
the characterization of dust in the ISM, there is still much debate 
about dust radiative transfer models and the appropriateness of  
techniques used in measuring $a$ and $g$ for a given sight line 
\citep{mat02, dra03, zub04}.  Further observations of dusty 
environments throughout our galaxy and beyond, along with the 
continued refinement of current dust models are needed in order 
to enhance our understanding.

The Upper Scorpius cloud is ideally suited for probing the scattering 
properties of dust grains.  It is an internally illuminated 
reflection nebula, which makes the total nebular flux strongly dependent 
on $a$ and only weakly dependent on $g$ \citep{gor94}.  This nebular geometry  
will result in an accurate measurement of $a$, but it is likely that any measure 
of $g$ will have large uncertainties.  The Upper Scorpius region 
was also found by \citet{gau93} to have very little foreground dust, which 
would add to the uncertainty in the scattering properties of the nebula.  
By considering this region as a whole, a large sample of UV bright stars 
are available as illumination sources for the dust model.  Multiple illuminating 
sources will reduce the model's dependence on the accuracy of any one
star's derived properties.  It was for these reasons that the nebula in 
Upper Scorpius was selected as the target for the Spectrograph for Photometric 
Imaging with Numeric Reconstruction (SPINR) sounding rocket mission (Cook et al. 2003, hereafter Paper I). 

The SPINR experiment provides a unique look at the ISM in the FUV wavelength 
regime from 912-1450 \AA.  By recording wide-field ($\sim$16\degr) 
spectral and spatial information from a given target simultaneously, a complete 
picture of the dust properties in a region can be derived.  Paper I
presents the details of the SPINR experiment while its data  
have been used by Lewis et al. (2005, hereafter Paper II) to derive the 
extinction properties along several lines of sight in the Upper Scorpius Cloud.
Paper II found that the Upper Scorpius region possesses a wide range of 
extinction properties, which makes it an excellent probe of the various dust 
environments seen throughout the Milky Way.  Thus far, only a handful of studies 
have probed dust scattering properties shortward of Ly$\alpha$ (1216 \AA) 
\citep{wit93, bur02, suj05, sha06, suj07}.  The SPINR data are unique in that spectra 
for the entire 16\degr\ field-of-view (FOV) were recorded simultaneously without the 
FOV limitations of a traditional long slit or the spectral resolution limitations of 
traditional wide band filters.  Currently, there are significant discrepancies 
between model predictions of dust albedo in the FUV and 
observations \citep{gor04}. The insights gained from the SPINR study of the 
Upper Scorpius cloud will help to paint a more complete picture of dust scattering 
properties in the FUV.      
     
The goal of this paper is to present the determination of $a$ and $g$ values 
for dust in the Upper Scorpius region at the effective wavelengths 
$\lambda_{eff}=$ 973 \AA, 1106 \AA, and 1261 \AA.  Additionally, the Upper 
Scorpius data presented in \citet{gor94} were revisited with an improved radiative 
transfer model to determine the value of $a$ in the STS-39 FUV camera (FUVCam) bandpasses, 
$\lambda_{eff}=$ 1362 \AA\ and 1769 \AA.  

\section{Observations and Image Reconstruction}\label{obs}

On 2000 August 4, the SPINR rocket experiment was launched from the 
White Sands Missile Range to record both
spectral and spatial information from the Upper Scorpius cloud in the FUV.
Approximately 400 s of data were recorded at a 2\arcmin.5
RMS spatial resolution and 10~\AA\ spectral resolution.  SPINR optically compresses 
the FOV into a single spatial dimension at each wavelength which in combination with the rotation of the rocket 
about its optical axis creates a three-dimensional (position, 
wavelength, and time) data cube. Extracting the stellar spectra 
from the SPINR data cube is a fairly simple operation.  One must simply 
identify the celestial coordinates corresponding to the axis of rotation 
of SPINR, then predict the phase and amplitude of the 
sine curve traced out by each star over the duration of the observation (Paper I).  
The point spread function (PSF) of the instrument varies with wavelength and position, but 
is on average Gaussian in shape with a full-width half-maximum of 10 pixels ($\sim$7\arcmin.8).  
Data lying within the PSF of the predicted sine curve can then be collected and assigned to 
the proper source.

\begin{figure}
\plotone{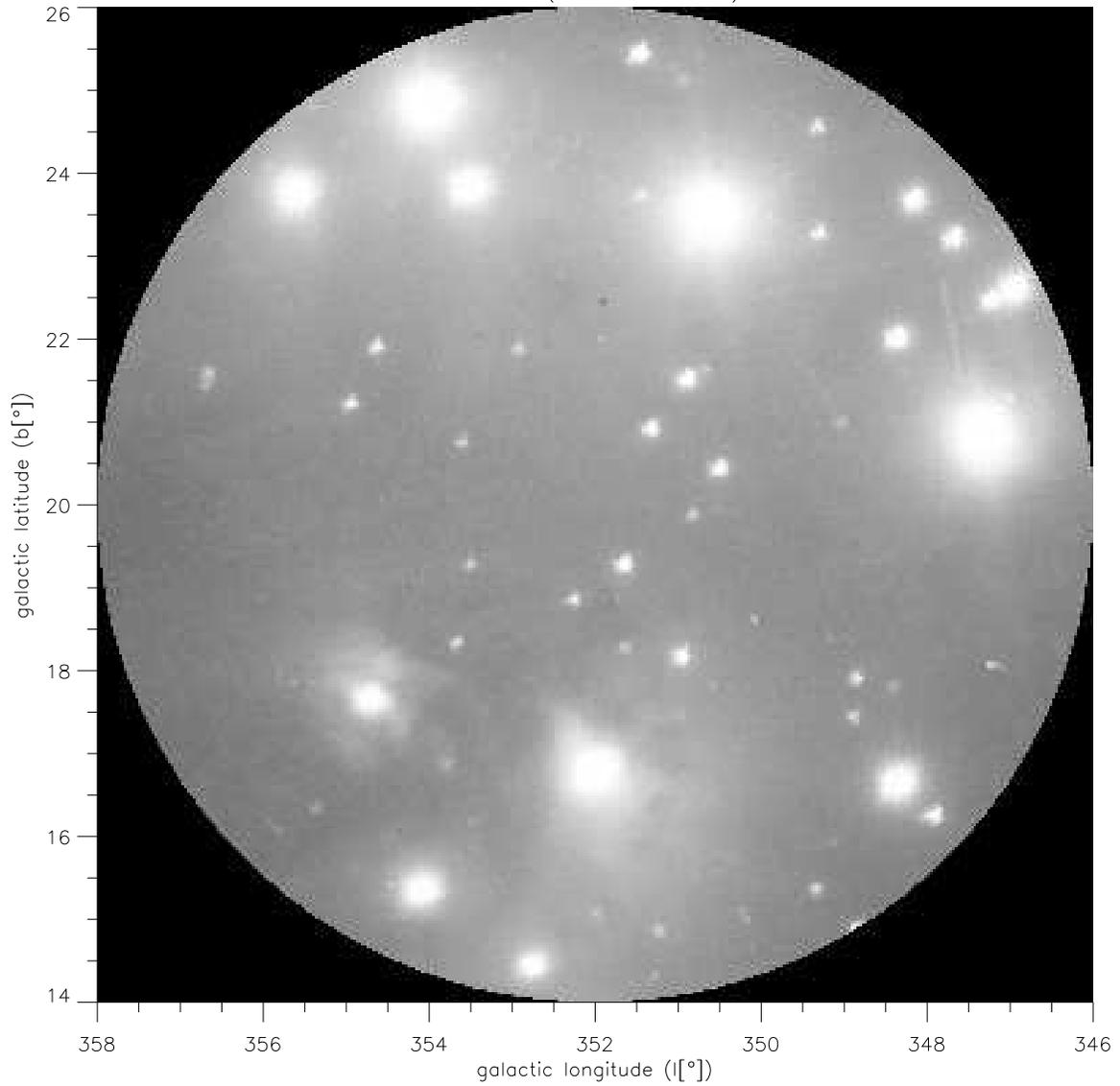}
\caption{Comparison of broadband (1230-1600 \AA) image taken with the STS-39 FUVCam (100s exposure)
and the reconstructed images in each of the SPINR wavelength bands.  Although the 
spatial resolution is lower in the SPINR images than the STS-39 FUVCam image, the basic structures of region
are clearly visible in the \longband\ bandpass image and in close correspondence 
with the STS-39 FUVCam image for 1230-1600 \AA.  The reconstructed SPINR images are included for visualization purposes only.  
All quantitative analysis is done using the raw SPINR sinogram data.}\label{recon}
\end{figure}

A similar process can be performed to 
reconstruct a spatial image of the SPINR data.  Paper I reviews 
several types of algorithms that could be used in the image reconstruction 
process.  In this study, the Pixon \citep{pin93} method of image reconstruction was used
to create images of the Upper Scorpius region in three arbitrary wavelength bands: 
\shortband, \midband, and \longband.  Figure \ref{recon} shows the 
reconstructed image for each SPINR wavelength bands and 
an image from the STS-39 FUVCam experiment (G. R. Carruthers 2006, private communication) 
for the 1230-1600 \AA\ wavelength range.  The FOV in each of these images has 
been constrained to the inner 10\degr\ of the the SPINR FOV for comparison purposes. 
The SPINR images are at a significantly lower spatial resolution than the raw SPINR data 
because of computational restrictions on the image reconstruction process.  For this reason, 
the data analysis presented here was performed on the raw SPINR sinogram data.  

\section{Data Analysis}\label{analysis}

Following the work of \citet{gor94}, there were several items that needed 
to be calculated in order to properly constrain the dust model.  One of the 
key items is the nebular to stellar flux ratio ($F_N/F_*$) for each 
wavelength band.  This ratio will constrain the range of acceptable solutions 
for the albedo, $a$.  The derivation of the total stellar flux, $F_*$, for 
each bandpass is detailed in section \ref{stellar}.  The basic properties 
of the stellar sources considered in this study are detailed in Table \ref{sd}.  
The derivation of the total nebular flux, $F_N$, is outlined in section \ref{neb}.
Table \ref{ratio} presents the values of $F_N/F_*$ for each bandpass derived in 
this study.  The model (further described in section \ref{model}) also requires 
estimates of the dereddened luminosity and the depth into the cloud of each 
stellar source, which are calculated by determining the extinction (section \ref{ext}), 
H$_{2}$ absorption (section \ref{h2abs}), and $R_V$ parameter (section \ref{tau}) 
along each line of sight.

\subsection{Dust Model}\label{model}
The DIRTY Monte Carlo radiative 
transfer model \citep{gor01, mis01} was used in this study to determine 
the $a$ and $g$ values that best reproduced the observed nebular surface brightness of the
Upper Scorpius cloud.  The geometric model of the Upper Scorpius cloud used is identical to
that of \citet{gor94} with a uniform sphere of dust with each
star embedded to a depth equal to its measured optical depth at each
wavelength modeled.  This simplistic geometry is used as it is
more important for the radiative transfer solution to have each star
embedded to the correct optical depth, than to the correct physical
depth.  The effects of the known clumpy structure of the ISM 
on albedo determinations \citet{mat02} are mitigated by the multiple 
illumination sources in the Upper Scorpius cloud.  The multiple
illumination sources mean that the computed scattered flux is relatively insensitive to the
details of a single star's parameters (luminosity, optical depth,
etc.), which could be affected by clumpy structures along its line of sight.  
This prescription for the radiative transfer model for Upper
Scorpius was shown to be valid, but not necessarily unique, by the close correspondence 
between the models and observed images in \citet{gor94}.

\subsection{Stellar Flux}\label{stellar}
The total stellar flux in each wavelength band was determined by identifying 
those stars in the SPINR FOV that are bright in the ultraviolet and simply 
summing their measured fluxes.  These UV bright stars were selected 
from the {\it Astronomical Netherlands Satellite} ({\it ANS}) catalog 
\citep{wes82} based on two criteria: 1) they reside inside the 16\degr\ SPINR FOV. 
2) at 1500 \AA\ the wide band ($\delta\lambda$=149 \AA) magnitude is less than 6.0 mag.  For the fourteen 
brightest stars, measurements of their reddened luminosities for each 
bandpass were easily extracted from the SPINR data by knowledge of their 
relative position.  Swaths of 30 pixels around the predicted stellar sine curves were 
extracted from the raw SPINR data in order to collect all the stellar light within $\sim6\sigma$ of the 
central Gaussian stellar peak.  Pixels where two or more stellar sine curves intersected were ignored.  
Those pixels not associated with a stellar line of sight were used to create a background spectrum 
that was removed from each of the stellar spectra.  This background spectrum includes the average 
nebular flux in the region in addition to the instrument background and telluric emissions.       
The flux from the dimmest nineteen stars in SPINR's FOV (15-33 in Table \ref{sd}) could not be disentangled 
from the nebular flux because of low spatial resolution and were estimated using available data and stellar models as described below.  
This flux amounts to less than 5\% of the total stellar flux in each bandpass.

For the \shortband\ and \midband\ bandpasses, estimates of the 
reddened luminosities for the dimmest nineteen stars were made using 
the Kurucz model \citep{kur92} corresponding to the surface gravity 
\citep{str81} and effective temperature \citep{dej87}
predicted from the MK spectral type and luminosity class of each star (Table \ref{sd}).  
The Kurucz models were then converted into SPINR units (counts $\rm s^{-1}$) using 
the dereddened spectrum of $\pi$ Sco as a standard.  However, in order to make 
an estimate of the expected reddened luminosity of each star, the amount of 
reddening along each line of sight had to be calculated.  Details on the derivation 
of extinction curves, H$_{2}$ attenuations, and ratios of total to selective 
extinction, $R_V$, can be found in the following sections.
Stellar fluxes derived from the Kurucz models 
accounted for less than 5\% of the total reddened stellar luminosity and less than 
10 \% of the total dereddened stellar luminosity. 

In the \longband\ bandpass, {\it International Ultraviolet Explorer} 
({\it IUE}) data (Table \ref{sd}) were converted into SPINR units 
(counts $\rm s^{-1}$) using the spectrum from $\pi$ Sco as a standard 
then summed over the bandpass to obtain an estimate of the reddened 
luminosity for the dimmest nineteen stars in the FOV.  
{\it IUE} data were also used to estimate the reddened luminosities 
of $\tau$ Sco and $\rho$ Sco in the \longband\ bandpass because their 
proximity to the edge of the detector prevented an accurate measurement in 
this lower sensitivity bandpass.  Table \ref{redlum} presents the reddened 
luminosities for each star that were used in calculating the total stellar 
flux ($F_*$) in each bandpass. 

\subsubsection{Extinction Curves}\label{ext}

The extinction curves of the form $k(\lambda-V)=E(\lambda-V)/E(B-V)$ for $\pi$ Sco, $\beta$ Sco, 
$\delta$ Sco, $\sigma$ Sco, $\omega$ Sco, $\nu$ Sco, and $\rho$ Oph, which account 
for $\sim$65\% of the total reddened stellar luminosity, were previously 
derived from SPINR data in Paper II. For the other stellar lines of sight 
in this study with lower signal-to-noise ratios, it proved difficult to derive a 
reliable measure of the extinction from the SPINR data in the 912 \AA\ to 1450 \AA\ 
wavelength range. However, as was shown in Paper II and \citet{sof05}, a fairly reliable prediction of extinction in the 
FUV along a given line of sight can be made by fitting a curve according to the parameterization 
of \citet{fit88} to {\it IUE} data then extrapolating the curve out shortward of Ly$\alpha$. 
For each line of sight possessing an $E(B-V)\geq 0.1$, an 
extinction curve was derived from the {\it IUE} data listed in Table \ref{sd} using the pair 
comparison method outlined in Paper II.  The comparison stars used and the subsequent 
\citet{fit88} parameterization for the extinction curves are found in Table \ref{FMCC}. For 
lines of sight with $E(B-V)<0.1$ it was assumed that the Galactic mean curve could be 
used to deredden the data with the introduction of only small errors [less than 1\% of 
$k(\lambda-V)$; \citep{gre92}].   

\subsubsection{H$_{2}$ Absorption}\label{h2abs}

When working with spectral data in the wavelength regime between the Lyman edge (912 \AA) 
and Ly$\alpha$ (1216 \AA) it is important to consider the effects of H$_{2}$ absorption.
In addition to the attenuation of starlight by dust in this wavelength regime, the H$_{2}$ 
Lyman and Werner band systems produce numerous absorption features in the spectra.
The H$_{2}$ optical depth templates developed by \citet{mcc03} for the {\it FUSE} mission were 
used in this study to make a best estimate of the H$_{2}$ attenuation, $\tau_{\rm H_2}$, along 
each line of sight.  \citet{sav77} presented measurements from the {\it Copernicus} spacecraft 
of the column densities of interstellar H$_{2}$ in the $J=0$ and $J=1$ rotational levels of the 
ground vibrational state ($v^{\prime\prime}=0$), $N(0)$ and $N(1)$ respectively.  Several 
of the lines of sight in this study have measurements of $N(0)$ and $N(1)$ detailed in 
\citet{sav77}.  For those lines of sight without readily available H$_{2}$ measurements, 
values for $N(0)$ and $N(1)$ were estimated from $E(B-V)$.

\citet{boh78} first derived the dust to gas ratio
\begin{equation}
N(\rm H_{tot})=(5.8\times10^{21}\mbox{ cm$^{-2}$ mag$^{-1}$})E(B-V)\label{htoteq},
\end{equation}
where $N(\rm H_{tot})=N(\hone)+2N(\rm H_2)$, from {\it Copernicus} data.
This relationship (equation \ref{htoteq}) was later confirmed by \citet{rac02} 
using {\it FUSE} data.  Additionally, \citet{dip94} used Ly$\alpha$ 
absorption measurements from {\it IUE} data to derived the ratio
\begin{equation}
N(\hone)=(4.93\pm0.28\times10^{21}\mbox{ cm$^{-2}$ mag$^{-1}$})E(B-V)\label{h1eq}.
\end{equation}
Combining equations (\ref{htoteq}) and (\ref{h1eq}), the total column density 
of molecular hydrogen along a given line of sight can be derived to 
be 
\begin{equation}
N(\rm H_{2})=(4.35\pm1.40\times10^{20}\mbox{ cm$^{-2}$ mag$^{-1}$})E(B-V)\label{h2eq}.
\end{equation} 
In this study it assumed that all of the interstellar $\rm H_2$ are 
in either the $J=0$ or $J=1$ rotational levels, therefore 
$N(\rm H_{2})=N(0)+N(1)$.  \citet{sav77} found an average value of 
the kinetic temperature of the gas to be $T_{01}=77\pm17$ K 
toward stars with $N(0)$ and $N(1)$ larger than $10^{18}$ $\rm cm^{-2}$.  
For the 11 lines of sight in this study with {\it Copernicus} measurements, all except $\tau$ Sco exhibited 
$N(0)$ and $N(1)>10^{18}$ $\rm cm^{-2}$.  Therefore, it assumed that $T_{01}=77\pm 17$ K, 
from which values for $N(0)$ and $N(1)$ can be derived using the 
Boltzmann relation and a value for $E(B-V)$.

With values for $N(0)$ and $N(1)$ for each line of sight estimated, the final  
item needed to produce an estimate $\tau_{\rm H_2}$ from the \citet{mcc03}
templates is an estimate of the Doppler parameter, $b$.  \citet{spi74} 
provided measurements of $b$ toward four stars in this study.  As shown 
in \citet{sav77}, with values for $N(0)\sim10^{19}$ $\rm cm^{-2}$, the use of 
estimated values of $b<10$ km $\rm s^{-1}$ should result in small ($\leq1\%$)
errors in the overall $\rm H_2$ profile.  Therefore, an average value 
of $b=5$ km $\rm s^{-1}$ was assumed toward each star, except for $\tau$ Sco 
where $N(0)\leq 1.66\times10^{13}$ $\rm cm^{-2}$, in which case the reported value of 
$b=3.8$ km $\rm s^{-1}$ was used.  Each of derived $\tau_{\rm H_2}$ templates 
were smoothed to the SPINR spectral resolution ($\sim10$\AA) then used in 
determining the total optical depth toward each star as described in 
section \ref{tau}.  

\subsubsection{$R_V$ Parameter and Optical Depths}\label{tau}

In this study, the ratio of total to selective extinction, $R_V=A(V)/E(B-V)$,
for each line of sight has been derived from available IR data 
according to the method presented in \citet{fit99}.  The majority of the IR data used in 
this study came from the Two Micron All Sky Survey (2MASS) database \citep{skr06}.  
However, the brightest stars in 
the SPINR FOV saturated the 2MASS instrument, so alternative IR measurements were sought from 
various sources.  For $\delta$ Sco (HD 143275), $\sigma$ Sco (HD 147165), $\rho$ Oph 
(HD 147933), and $\chi$ Oph, J(1.25 \micron), H(1.65 \micron), K(2.2 \micron) band data 
were taken from \citet{the86}.  Additionally, IR data from \citet{lei84}, \citet{car90}, 
and \citet{joh66} were used for determining $R_V$ along the lines of sight to 
$\beta$ Sco, $\omega$ Sco, and $\nu$ Sco respectively.  

Values for $E(V-J)$, $E(V-H)$, and $E(V-K)$ were then calculated using the intrinsic 
color indices from \citet{weg94} corresponding to the MK classifications 
presented in Table \ref{sd}.  For lines of sight with $E(B-V)<0.1$, $R_V=3.1$ 
was assumed since these small values would generally lead to large discrepancies 
in the calculation of $R_V$.  Because $\alpha$ Sco B (HD 148479) is the companion 
star to an IR bright M type star (HD 148478), it proved difficult to use IR data 
to predict the value of $R_V$ in the $\alpha$ Sco region, therefore
the $R_V$ value of its nearest neighbor (HD 148605) was used.  The derived values 
for $R_V$ are presented in Table \ref{tautab}.

The optical depth as a function of wavelength, $\tau(\lambda)$, for each line 
of sight was calculated by
\begin{equation}
\tau(\lambda)=\frac{E(B-V)}{1.086}[k(\lambda-V)+R_V]+\tau_{\rm H_2}(\lambda),
\end{equation} 
using the derived extinction curves ($k(\lambda-V)$), $\rm H_2$ attenuations
($\tau_{\rm H_2}(\lambda)$), and $R_V$ parameters.  With $\tau(\lambda)$ 
determined, the dereddened luminosity of each star in the FOV is simply
calculated by
\begin{equation}
L_0(\lambda)=L(\lambda)e^{\tau(\lambda)},
\end{equation}
where $L_0$ represents the dereddened luminosity and $L$ represents the 
reddened luminosity.  The $L_0$ spectra
were then summed over the appropriate wavelength bands (Table \ref{deredlum}). 

The effective optical depth, $\tau_{eff}$, was calculated using the 
weighted average over each of the SPINR wavelength bands given by
\begin{equation}
\tau_{eff}=\frac{\sum_{\lambda_1}^{\lambda_2}\tau(\lambda)L(\lambda)}
                {\sum_{\lambda_1}^{\lambda_2} L(\lambda)},  
\end{equation} 
where $L(\lambda)$ includes SPINR's sensitivity function.  Overall,
the contributions of H$_{2}$ absorption, $\tau_{\rm H_2}$, to $\tau_{eff}$ 
in each wavelength band are small compared to extinction by dust [$\sim$0\% in the 
\longband\ band, $\sim$6\% in \midband\ band, and $\sim$14\% in the \shortband\ band].  
     
\subsection{Nebular Flux}\label{neb}
\begin{figure}
\plotone{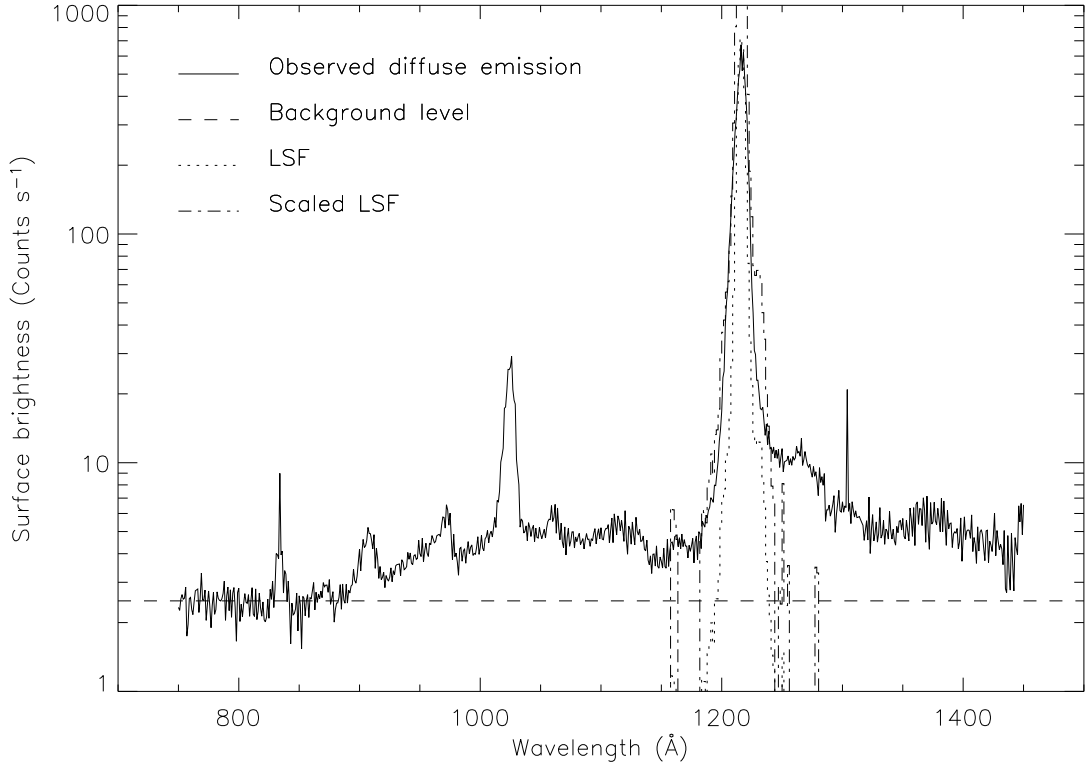}
\caption{The raw observed diffuse brightness and the derived background level.  
In addition to the background level (which is a combination of the
detector dark rate and scattered Ly$\alpha$) this figure shows the
approximate Ly$\alpha$ LSF.  The LSF is from
calibration data shifted in wavelength and scaled to match the observed
emission line strength.  A second LSF is shown scaled to the strength
appropriate for the wings of Ly$\alpha$ as they impinge on the high
sensitivity areas of the detector (away from the core of Ly$\alpha$).
For further details of the detector sensitivity see Paper I.}\label{bglsfplot}
\end{figure}

The total nebular flux was determined by taking an average over the detector 
pixels that were not known to be part of a stellar spectrum.  Due to the nature of the experiment's geometry 
these nebular pixels are not specific to any one portion to the Upper Scorpius Cloud, but instead probe 
the nebulosity over the entire SPINR FOV.  The radiative transfer model assumes a 
uniform sphere of dust, which best measures the average nebulosity in the Scorpius 
region as constrained by the average nebular flux. Figure \ref{bglsfplot} shows 
the unprocessed spectrum extracted for the nebular flux.  An estimate of the 
background was determined from the mean of the spectrum from 750-800 \AA\, which should 
be essentially equal to zero.  It is easy to see 
that this spectrum is contaminated with telluric emission and may contain emission 
from other sources besides dust such as H$_{2}$ fluorescence.  The most prominent 
telluric lines in the nebular spectrum are from \mbox{\ion{O}{2}} at 834 \AA, 
a collection of higher lyman lines and lyman continuum from 906 to 918 \AA, 
\hone\ at 972 \AA\ (Ly$\gamma$), 1025 \AA\ (Ly$\beta$),
and 1216 \AA\ (Ly$\alpha$), and  \mbox{\ion{O}{1}} at 1304 \AA\ \citep{cha84}.  These telluric 
emissions were corrected for by determining the continuum level under each feature 
and removing the excess flux in those regions.  The asymmetry in the Ly$\alpha$ feature is the 
result of the wings of the Ly$\alpha$ line impinging on the higher sensitivity areas of the detector (Paper I).  
Figure \ref{bglsfplot} presents the Ly$\alpha$ line spread function (LSF) scaled to account 
for this asymmetry. This scaled LSF was used to determine the appropriate amount of flux to attribute 
to Ly$\alpha$ in the \longband\ bandpass.

\begin{figure}
\plotone{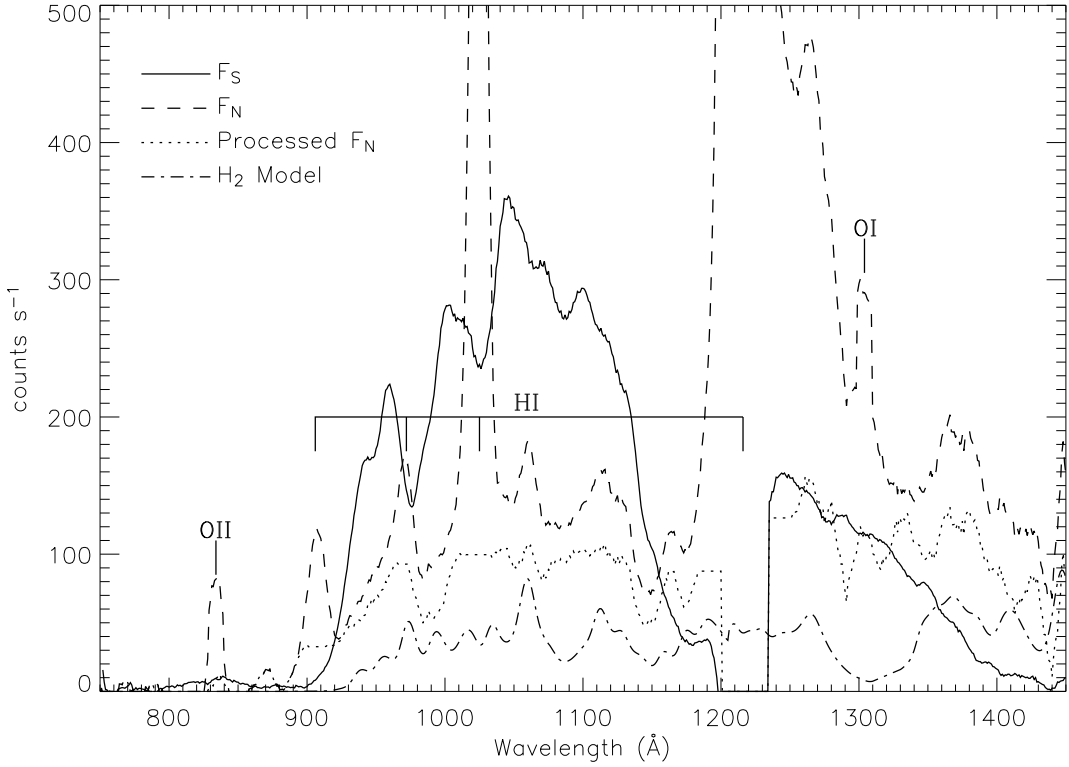}
\caption{Comparison of the nebular (dotted line) and stellar (solid line) spectra 
derived from SPINR data.  Telluric lines that were removed from the unprocessed nebular (dashed line)
spectrum are noted.  Also, the modeled H$_{2}$ emission removed from nebular spectrum 
is presented (dash-dot line).  The nebular spectrum remains relatively constant across
all bandpasses, while the stellar spectrum peaks in the \midband\ bandpass.  This directly 
corresponds to derived nebular to stellar flux ratios (Table \ref{ratio}), which 
shows a steady increase in the ratio from the \shortband\ to the \longband\ bandpass.}\label{ratplot}
\end{figure}

Determining how much, if any, H$_{2}$ fluorescence was contributing to the nebular flux 
involved producing a H$_{2}$ emission model that best fit the data as described in section \ref{h2em}.  The 
final value of the nebular flux, $F_N$, for each bandpass was determined by 
\begin{equation}
F_N=\sum_{\lambda_1}^{\lambda_2}I_N(\lambda)-\sum_{\lambda_1}^{\lambda_2}I_{H_{2}}(\lambda)
-\sum_{\lambda_1}^{\lambda_2}I_{telluric}(\lambda),
\end{equation}
where $I_N(\lambda)$ represents the unprocessed nebular spectrum, 
$I_{\rm H_{2}}(\lambda)$ is the H$_{2}$ fluorescence model, and $I_{telluric}(\lambda)$ 
represents identified telluric emission.  Additionally, a small correction (less than 5\%) was subtracted from the nebular flux
to account for any residual starlight that may have been lumped in with the nebular spectrum.  Uncertainties 
in the determination of $I_{telluric}(\lambda)$ and $I_{\rm H_{2}}(\lambda)$ were propagated to determine 
the overall uncertainty in $F_N$ and hence $F_N/F_*$ (Table \ref{ratio}).  Figure \ref{ratplot} 
presents the processed nebular spectrum, which has been scaled to match the appropriate detector area.

\subsubsection{H$_2$ Emission}\label{h2em}
Molecular hydrogen (H$_{2}$) is the most abundant molecule in the 
ISM, with dipole-allowed band systems that span $\sim$912-1650 \AA.
These are readily observed as a series of strong absorption features 
in the 912-1110 \AA\ spectra of hot stars (as discussed in section \ref{h2abs}).  
The re-emission that follows as the molecules cascade back to excited
rovibrational levels of the ground electronic state 
produces a highly structured spectrum that has been observed in the diffuse 
ISM \citep{mar90,lee06} and in photodissociation regions
near massive stars \citep{wit89,fra04,fra05b}.  
Consequently, efforts to determine the grain properties from scattered light 
observations at FUV wavelengths should account for the 
contribution, if any, from H$_{2}$ fluorescence \citep{wit93,sha06}.

This study used a model of H$_{2}$ fluorescence to assess the importance
of these emissions and correct for their contribution to the nebular
spectra obtained by the SPINR experiment.  The 
H$_{2}$ fluorescence model, described in \citet{fra05a}, was adopted 
for the present work,  which takes input parameters such as the
total column density, Doppler $b$-value, and the distribution of molecules
among the rovibrational levels of the ground electronic state.
The code computes photoexcitation rates into the upper electronic states
that depend on the strength of the illuminating radiation field
at the transition wavelength.  These excited levels 
decay to the ground electronic state, obeying the appropriate branching
ratios and selection rules \citep{abg93a,abg93b}.  These transitions
produce the synthetic fluorescence spectrum.

The unprocessed nebular spectrum presented in Figure \ref{ratplot} shows the presence of 
emission features in addition to those attributed to telluric lines, which is an indication 
of the likely presence of H$_{2}$ fluorescence in the Upper Scorpius region.
However, the FOV and spectral resolution of the SPINR instrument 
makes a determination of the exact properties of the fluorescent
H$_{2}$ challenging, thus a template spectrum was created
that can be scaled to the observed flux level, enabling the 
subtraction of the flux due to H$_{2}$ fluorescence from the nebular data.  
Typical values for the column density of the diffuse ISM
(N(H$_{2}$)~=~10$^{20}$ cm$^{-2}$) and $b$-value (2 km s$^{-1}$)
were assumed for the model.  The ground state population distribution
is uncertain.  Noting that several studies of 
photoexcited H$_{2}$ emission are characterized by rotational 
excitation temperatures of order several hundred K \citep{mar99,hab04,all05,fra07}, 
A value of T(H$_{2}$) = 500 K was adopted.
By contrast, observational studies suggesting 
radiatively excited H$_{2}$ at the diffuse ISM H$_{2}$ kinetic temperature of 
$\sim$75 K \citep{sav77,bur07} have not been found in the ISM.
It is likely that within the large SPINR FOV cold (~77 K) H$_{2}$ populations that are 
responsible for the absorption of stellar UV light coexist with warm 
(~500 K) H$_{2}$ populations that are responsible for the observed H$_{2}$ fluorescence. 

The illuminating radiation field was assumed to be a lightly reddened B0IV star, 
characteristic of the spectral type of the hot stars in
Scorpius and the average interstellar radiation field \citep{dra78}.
This radiation field was created from $FUSE$ spectra of a
B0IV star from 916.6-1181.9 \AA, covering essentially
all of the absorbing transitions for T(H$_{2}$) = 500 K.
Since the template spectrum was scaled, changes in absolute flux
were not of concern. Changes in spectral shape caused by varying
the temperature and radiation field were largely washed out
by convolving the model to the 10~\AA\ instrumental resolution.

In summing over the large SPINR FOV, an ensemble of 
illuminating radiation fields and H$_{2}$ column densities were taken, 
thus no attempt was made to reproduce the absolute flux of the observed emission.
The individual rotational lines are not resolved with SPINR, and the long wavelength 
cut-off (1450 \AA) does not permit a measurement of the 1578/1608 ratio. However, 
detection of H$_{2}$ fluorescence and dust scattered light with SPINR
allows the determination of their relative contribution to the FUV 
light from Scorpius.  The scaled synthetic H$_{2}$ fluorescence
spectrum and the residual dust scattered light from the 
SPINR nebular data were integrated over three of the model component bands 
(\shortband, \midband, \longband).  The relative contribution of H$_{2}$ emission to the 
diffuse FUV light in Scorpius in each of the SPINR bandpasses is defined as 
\begin{equation}
f^{H_{2}}_{j}~=~\frac{\sum_{\lambda_1}^{\lambda_2} I_{H_{2}}(\lambda)}
{\sum_{\lambda_1}^{\lambda_2}(F_N(\lambda) +  I_{H_{2}}(\lambda))},
\end{equation} 
where $j$ is the FUV bandpass, $I_{\rm H_{2}}(\lambda)$ is the fluorescence model, 
and $F_N(\lambda)$ is the dust scattered light.
The relative contribution of H$_{2}$ fluorescence to the nebular light
is $\approx$~29\% in all three bands, with individual values of
[$f^{H_{2}}_{912-1029}$, $f^{H_{2}}_{1030-1200}$, $f^{H_{2}}_{1235-1450}$] =
[0.272~$\pm$~0.069, 0.314~$\pm$~0.079, 0.284~$\pm$~0.071].     

This determination of $f^{H_{2}}_{j}$ is consistent with early
theoretical considerations by~\citet{jak82}.  Using
 bandpass parameters similar to those chosen for the SPINR observations, 
he predicts that H$_{2}$ fluorescence could account for $\sim$30\% of the total nebular flux
($f^{H_{2}}_{1500}$~$\sim$~0.3).
This finding is also in agreement with the Berkeley UVX Shuttle Spectrometer 
observations of the diffuse ISM~\citep{mar90}, which provided the first
direct detection of H$_{2}$ fluorescence in the diffuse ISM.
The SPINR results are in  excellent agreement
with their Targets 4 ($l$,$b$=168\arcdeg,-16\arcdeg; 
$f^{H_{2}}_{1550}$~=~0.27) and 6 (142,35; $f^{H_{2}}_{1550}$~=~0.32), 
and reasonable agreement with their Targets 
2 (132,40; $f^{H_{2}}_{1550}$~=~0.33) and
5 (135,25; $f^{H_{2}}_{1550}$~=~0.25).

\section{Results and Discussion}\label{dis}

\begin{figure}
\plotone{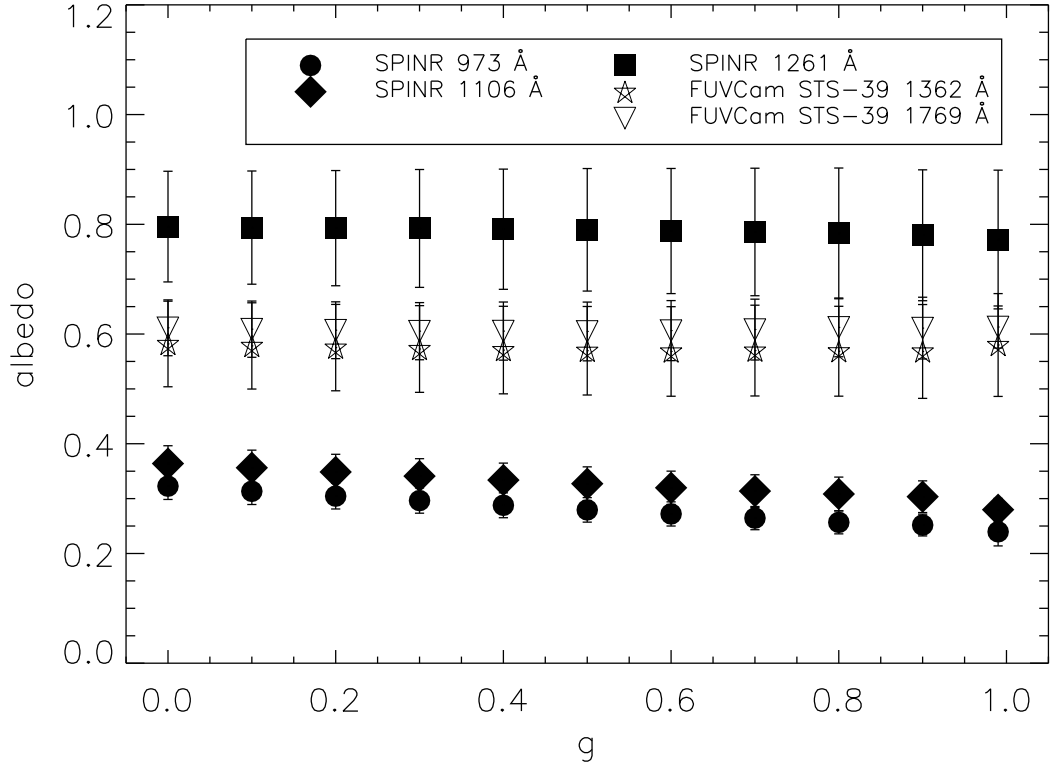}
\caption{Optimal dust albedo values as a function of $g$ for the SPINR and STS-39 FUVCam data. As can be seen there
is very little variation in the derived values of $a$ across all values of $g$.  This is a direct result of the 
spherical cloud with embedded stellar sources geometry of the Upper Scorpius region.}\label{ag} 
\end{figure}

Dust model simulations were run for each of the SPINR wavelength bands 
to identify the value for the dust albedo, $a$, that produced the best 
fit to the data.  This was accomplished by finding the $a$ value that 
reproduced the $F_N/F_*$ for each bandpass while varying the asymmetry 
parameter, $g$, from $0.0$ to $0.99$.  As shown in Figure \ref{ag}, 
there was very little variation in the best-fit value for the albedo over the 
entire range of $g$ values for all three of the SPINR bandpasses and the 
two STS-39 FUVCam bandpasses.  The best fit values for the dust 
albedo were $a=0.28\pm0.07$ in the \shortband\ bandpass, $a=0.33\pm0.07$ in 
the \midband\ bandpass, and $a=0.77\pm0.13$ in the \longband\ bandpass.  
Additionally, the STS-39 data presented in \citet{gor94} were used in 
the updated radiative transfer model to make new estimates of the dust
albedo in the Upper Scorpius region for the STS-39 FUVCam 1230-1600 \AA 
1650-2000 \AA bandpasses.  The new albedo values 
that provided the best fit to the STS-39 FUVCam data were slightly modified
from the original estimate of $a=0.47-0.70$ for $\lambda_{eff}=1362 \AA$ 
and $a=0.55-0.72$ for $\lambda_{eff}=1769 \AA$, to $a=0.54\pm0.06$ and 
$a=0.59\pm0.07$ respectively.

The optical depth of each star in the model Scorpius cloud is determined 
mostly from the derived extinction curves, but also has some dependence 
on the values selected for $R_V$ and $\tau_{\rm H_2}$ (see section \ref{tau}).  
In order to test the strength of the dependence of the derived albedo 
values on $R_V$ and $\tau_{\rm H_2}$, simulations were run for models 
with no H$_2$ absorption and $R_V=3.1$ for every stellar line of sight.  
Not accounting for H$_2$ absorption and using a uniform value of $R_V=3.1$ 
reduced the albedo values by $<$5\%, which is still well within the 
quoted uncertainties for $a$.  Dust model simulations were also run 
where presence of H$_2$ fluorescence in the Upper Scorpius cloud was ignored
(see section \ref{h2em}). Not accounting for H$_2$ emission in the nebular 
data results in a $\sim$40\% increase in $F_N/F_*$.  This increase in $F_N/F_*$ 
resulted in values for the albedo of $a=0.35\pm0.05$ in the \shortband\ bandpass, 
$a=0.43\pm0.05$ in the \midband\ bandpass, and $a=0.93\pm0.04$ in the \longband\ 
bandpass, which are all within 2$\sigma$ of the best albedo estimates accounting for
H$_2$ emission.  Overall, the derived albedo values are not particularly sensitive to the 
amount of H$_2$ absorption or $R_V$ value selected along any given line 
of sight.  However, moderate inconsistencies can be introduced into the albedo values 
if H$_2$ fluorescence is not considered.
  
Because SPINR was able to record spatial information 
about the dust in the Upper Scorpius region in addition to spectral 
information, it was possible to determine the value of $g$ that best 
reproduced the SPINR data.  Model images were produced over a range of $g$ values 
from 0.00 to 0.99 and with $a$ values corresponding to the best fit for each
$g$ value (Figure \ref{ag}). These model images were ``spun'' into sinogram 
space using a transformation matrix and then compared to the SPINR 
sinograms using a cross-correlation metric.  The error in the cross-correlation 
metric was determined using a Monte Carlo simulation in which up to $\pm 1\sigma$ 
worth of noise was added to each pixel in the SPINR sinogram.  The optimal $g$ value was 
determined from the image/sinogram with the highest cross-correlation to the 
SPINR data.  The range of acceptable $g$ values was determined from the Monte Carlo 
error estimate on the highest cross-correlation.  All $g$ cross-correlation metrics 
that fell within the error of the optimal fit were deemed acceptable.

\begin{figure}
\plotone{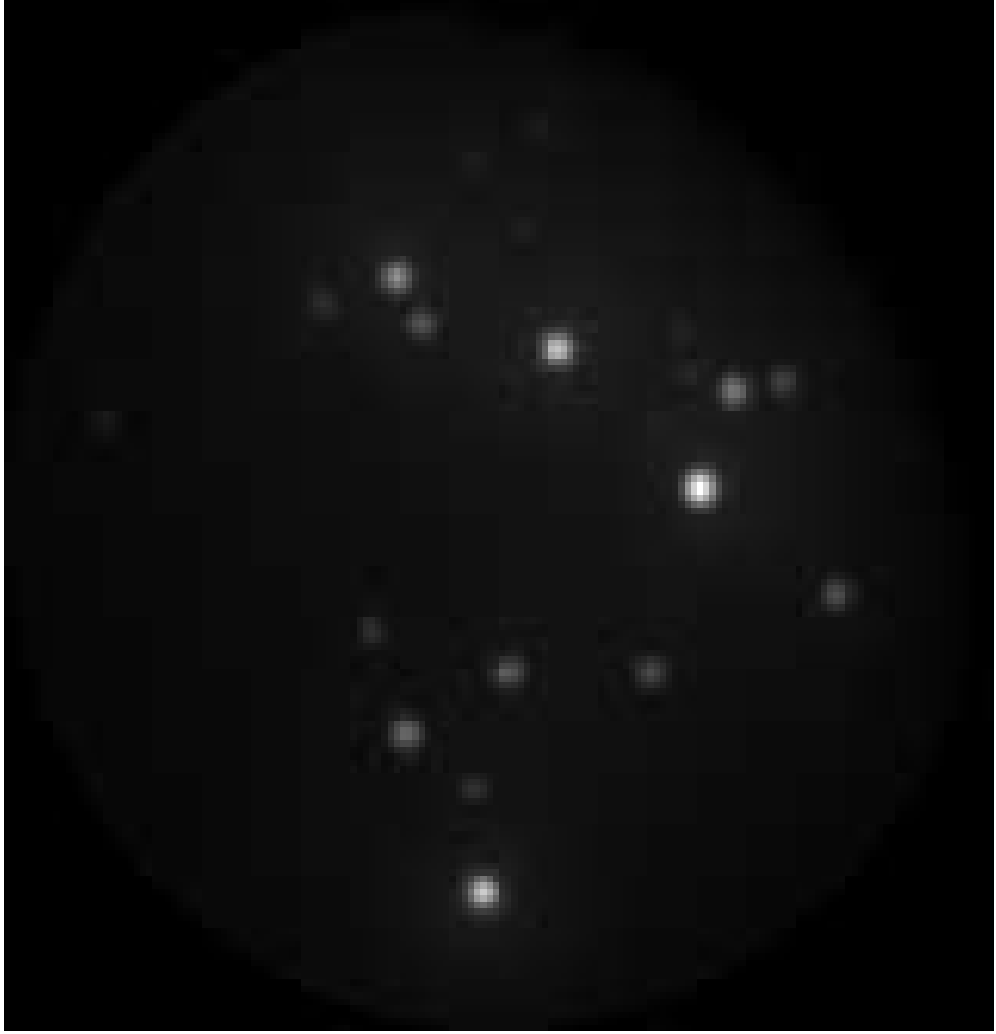}
\caption{Comparison of the SPINR data, best fit dust model, and a model outside the acceptable ranges for $a$ and $g$ 
for the \midband\ bandpass.  The cross-correlation metric between the sinograms was used in determining the optimal 
$g$ value and the uncertainty in this measurement for the \midband\ bandpass.  Model/observation comparisons were made 
in sinogram space because the transformation from image to sinogram space is more precisely known than the transformation 
from sinogram to image space.  The signal-to-noise ratio is higher for the model, which accounts for the small 
differences seen between the images.  The $a=0.75$ and $g=0.0$ model is brighter in the regions between the 
stellar sources, which is not generally seen in the SPINR observations.}\label{sino}
\end{figure}

The $g$ value that provided the best fit to the SPINR \midband\ bandpass data was 
$g=0.96^{+0.02}_{-0.18}$.  Figure \ref{sino} shows sinogram and image data for the SPINR 
\midband\ bandpass, the corresponding best fit dust model($g=0.96$ and $a=0.30$), and a 
model outside of the range of acceptable $a$ and $g$ values ($g=0.0$ and $a=0.75$).  In the 
\shortband\ and \longband\ bandpasses no statistically significant constraints could 
be put on the $g$ value.  It is not surprising that values for $g$ are not well determined in this 
study since the geometry of the nebula in Upper Scorpius was selected for its weak dependence on 
$g$ to better determine $a$.  However, given the constraint from the \midband\ bandpass 
we can confidently say that $g>0.42$ at the three sigma level.   This is not a tight 
constraint on $g$, but does align with the general expectation for $g$ to be highly forward scattering, 
$g>0.7$, as shown by previous predictions for $g$ in the FUV \citep{gor04}.

\begin{figure}
\plotone{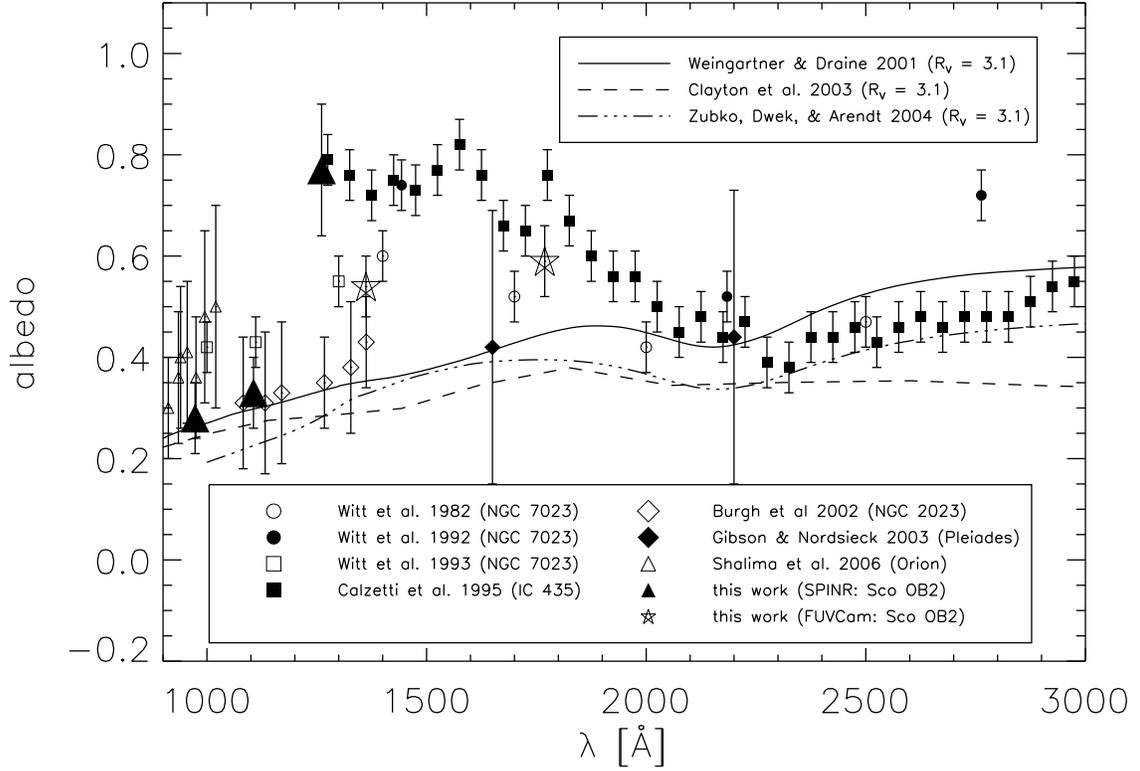}
\caption{A plot of albedo ($a$) vs. wavelength for the values derived in this study 
along with other values from literature for reflection nebulae.  The values for 
the albedo derived in this study show strong correspondence with albedo values 
derived from other observational studies.  The derived albedos for $\lambda_{eff}=973$ \AA\ 
and $1106$ \AA\ are well correlated with the results from \citet{cal95} and 
the dust model of \citet{wei01}.  The derived albedo for $\lambda_{eff}=1261$ \AA\ is 
higher than would nominally be expected from the STS-39 FUVCam observations, but still within 1$\sigma$ of other albedo 
measurements in this spectral region, which all show significant deviation from model predictions.}\label{alit}
\end{figure}

The results for $a$ from this study are compared with 
previous studies of reflection nebulae in Figure \ref{alit}. The values of $a$ 
derived for the \shortband\ and \midband\ bandpasses correspond well with both
previous observational determinations $a$ and the \citet{wei01} model in this 
spectral region.  The value of $a$ determined for the \longband\ bandpass is on the 
high end of the observational data points scattered within the 1200-1400 \AA\ range, 
but are well within $3\sigma$ of the other observational determination of $a$.  However, 
what is clear from this study and previous observational studies of albedo in the FUV 
is that the dust models \citep{wei01,cla03,zub04} do not correctly account for the apparent 
rise in the albedo from around 1200 \AA\ to 1800 \AA.  Further observational studies of the 
ISM in this spectral region are needed to fully characterize this rise in the albedo 
and determine its source.

\section{Summary}\label{sum}
Both the stellar and nebular populations of the Upper Scorpius region 
have been characterized in the FUV (912-1450 \AA) in this study.  The 
derived nebular and stellar flux levels were used to constrain a 
radiative transfer model and make a determination of the $a$ and $g$ values that
the dust grains in the region possess for three bandpasses (\shortband, \midband, and \longband).
Overall, the values for $a$ and $g$ derived in this study corroborate 
previous findings that dust in the ISM is highly forward scattering
($g>0.7$) and possess a low albedo ($a<0.4$) in the 912-1200 \AA\ 
wavelength regime.  

\acknowledgments
The authors would like to thank Dr. George S. Carruthers for providing us 
with the images of the Upper Scorpius region from the STS-39 FUVCam 
experiment, the Wallops sounding rocket staff for their support, and 
Ms. Valerie Gsell for directing the payload. We would also like to thank our 
anonymous referee for their helpful comments and suggestions.
This publication makes use of data products from the Two Micron All Sky Survey, 
which is a joint project of the University of Massachusetts and the Infrared 
Processing and Analysis Center/California Institute of Technology, funded by 
the National Aeronautics and Space Administration and the National Science 
Foundation.  
This work was supported in part by NASA grant NAG5-690.
\appendix

\bibliographystyle{apj}
\bibliography{ms_astroph}

\clearpage

\begin{deluxetable}{ccccccccc}
\tablecolumns{9}
\tablewidth{0pt}
\tablecaption{Stellar Data \label{sd}}
\tablehead{\colhead{No.} &
\colhead{HD} & \colhead{Name} & \colhead{l(\degr)} & \colhead{b(\degr)} & \colhead{MK\tablenotemark{a}} &
\colhead{V$_{mag}$\tablenotemark{a}} & \colhead{E(B-V)\tablenotemark{b}} & \colhead{{\it IUE} Data}}
\startdata
1 & 149438 & $\tau$ Sco & 351.535 & 12.807 & B0V & 2.82 & 0.034 & swp33008\\
2 & 143275 & $\delta$ Sco & 350.096 & 22.491 & B0.3IV & 2.32 & 0.120 & ...\\
3 & 144217 & $\beta$ Sco & 353.197 & 23.601 & B1V & 2.62 & 0.170 & ...\\
4 & 143018 & $\pi$ Sco & 347.214 & 20.231 & B1V & 2.89 & 0.065 & ...\\
5 & 147165 & $\sigma$ Sco & 351.314 & 16.998 & B1III & 2.89 & 0.340 & ...\\
6 & 142669 & $\rho$ Sco & 344.628 & 18.272 & B2IV-V & 3.88 & 0.022 & swp32877\\
7 & 144470 & $\omega$ Sco & 352.751 & 22.772 & B1V & 3.96 & 0.190 & ...\\
8 & 145502 & $\nu$ Sco & 354.608 & 22.700 & B3V & 4.01 & 0.250 & ...\\
9 & 145482 & 13 Sco & 348.117 & 16.836 & B2V  & 4.59 & 0.044 & ...\\
10 & 148605 & 22 Sco & 353.099 & 15.796 & B2V  & 4.79 & 0.069 & swp09221\&lwr07977\\
11 & 141637 & 1 Sco & 346.098 & 21.705 & B3V & 4.64 & 0.160 & swp42216\&lwp20988\\
12 & 142114 & 2 Sco & 346.876 & 21.614 & B2.5Vn & 4.59 & 0.140 & swp20611\&lwr16524\\
13 & 147933 & $\rho$ Oph & 353.687 & 17.687 & B2IV & 4.63 & 0.450 & ...\\
14 & 148479\tablenotemark{c} & $\alpha$ Sco B & 351.946 & 15.065 & B3V & 4.91 & 0.150 & swp05892\&lwr05146\\
\hline
15 & 148184 & $\chi$ Oph & 357.932 & 20.677 & B2IVpe & 4.42 & 0.490 & swp54090\&lwp30205\\
16 & 142096 & $\lambda$ Lib & 350.725 & 25.378 & B2.5V & 5.03 & 0.200 & swp42326\&lwr10778\\
17 & 142184 & ... & 347.932 & 22.545 & B2.5Vne & 5.42 & 0.155 & swp36741\&lwp15993\\
18 & 142990 & ... & 348.121 & 21.196 & B5IV & 5.43 & 0.090 & swp42227\\
19 & 142165 & ... & 347.514 & 22.149 & B6IVn & 5.39 & 0.113 & swp42217\&lwp17717\\
20 & 142378 & 47 Lib & 351.648 & 25.657 & B5V & 5.94 & 0.140 & swp09236\&lwr07996\\
21 & 142883 & ... & 350.885 & 24.085 & B3V & 5.85 & 0.200 & swp36751\&lwp16004\\
22 & 144334 & ... & 350.348 & 20.855 & B8p & 5.92 & 0.081 & swp21083\&lwr16819\\
23 & 142301 & 3 Sco & 347.124 & 21.513 & BB8IIIp & 5.87 & 0.110 & swp21092\&lwr16825\\
24 & 142250 & ... & 345.568 & 20.005 & B6Vp & 6.14 & 0.050 & swp36742\\
25 & 146001 & ... & 350.386 & 18.119 & B7IV & 6.05 & 0.153 & swp38567\&lwp16003\\
26 & 145792 & ... & 351.011 & 19.029 & B5V & 6.41 & 0.169 & swp36749\&lwp16002\\
27 & 144661 & ... & 349.992 & 19.970 & B7IIIp & 6.33 & 0.097 & swp13953\\
28 & 145483 & 12 Sco & 347.745 & 16.499 & B9V & 5.67 & 0.059 & swp16306\\
29 & 144844 & ... & 350.734 & 20.369 & B9IVp & 5.88 & 0.116 & swp36828\\
30 & 142884 & ... & 348.963 & 22.255 & B9p & 6.79 & 0.159 & swp36743\\
31 & 146416 & ... & 353.981 & 20.596 & B9V & 6.61 & 0.075 & swp36831\\
32 & 142315 & ... & 348.980 & 23.299 & B8V & 6.86 & 0.116 & swp36837\&lwp16118\\
33 & 145102 & ... & 348.546 & 17.872 & B9Vp & 6.59 & 0.144 & swp36829\&lwp16110\\
\hline
\multicolumn{7}{l}{Additional Standard Stars\tablenotemark{d}}\\
\hline
... & 122980 & $\chi$ Cen & 317.730 & 19.538 & B2V & 4.35 & 0.04 & swp46857\&lwp24818\\
... & 32630 & $\eta$ Aur & 165.354 & 00.272 & B3V & 3.17 & 0.02 & swp08197\&lwr07125\\
... & 3360 & $\zeta$ Cas & 120.776 & -08.914 & B2IV & 3.66 & 0.04 & swp04316\&lwr03812\\
... & 61831 & ... & 252.138 & -07.898 & B2.5V & 4.84 & 0.03 & swp14309\&lwr10940\\
... & 90994 & $\beta$ Sex & 246.415 & 46.167 & B6V & 5.07 & 0.00 & swp15791\&lwr12162\\
... & 34759 & $\rho$ Aur & 166.565 & 02.927 & B5V & 5.22 & 0.01 & swp15537\&lwr09868\\
... & 29335 & 49 Eri & 195.134 & -28.939 & B7V & 5.31 & 0.00 & swp15788\&lwr12159\\
... & 222173 & $\kappa$ And & 109.766 & -16.714 & B9IVn & 4.14 & 0.07 & swp33853\&lwp13557\\
... & 196867 & $\alpha$ Del & 060.304 & -15.321 & B9V & 3.77 & 0.01 & swp15545\&lwr12025
\enddata
\tablenotetext{a}{From \citet{hof91}.}
\tablenotetext{b}{E(B-V) values taken preferentally from \citet{pap91} then
\citet{deg89} and finally \citet{weg03}.}
\tablenotetext{c}{MK, V$_{mag}$, and E(B-V) for HD 148479 were taken from \citet{sno87}.}
\tablenotetext{d}{All standard star data taken from \citet{wu98}.}
\end{deluxetable}

\clearpage

\begin{deluxetable}{lcc}
\tablecolumns{3}
\tablewidth{0pt}
\tablecaption{Nebular to Stellar Flux Ratios\label{ratio}}
\tablehead{\colhead{Bandpass} & \colhead{$\lambda_{eff}$} & \colhead{$F_N/F_*$}}
\startdata 
912\AA-1029\AA & 973\AA & 0.377$\pm$0.041 \\
1030\AA-1200\AA & 1106\AA & 0.406$\pm$0.052\\     
1235\AA-1450\AA & 1261\AA & 1.301$\pm$0.159 
\enddata 
\end{deluxetable}

\clearpage

\begin{deluxetable}{lccc}
\tablecolumns{4}
\tablewidth{0pt} 
\tablecaption{Measured Reddened Luminosities\label{redlum}} 
\tablehead{ \colhead{HD} & \colhead{$L$[973\AA]} & \colhead{$L$[1106\AA]} & \colhead{$L$[1261\AA]}\\
\colhead{} & \colhead{(counts $\rm s^{-1}$ $\rm bandpass^{-1}$)} & 
\colhead{(counts $\rm s^{-1}$ $\rm bandpass^{-1}$)} & 
\colhead{(counts $\rm s^{-1}$ $\rm bandpass^{-1}$)}}
\startdata
149438 & 6461.77$\pm$42.01 & 7068.90$\pm$39.63 & 3620.92$\pm$715.02\\
143275 & 4592.08$\pm$26.21 & 7523.04$\pm$32.92 & 3360.57$\pm$27.56 \\
144217 & 2319.94$\pm$21.79 & 4691.78$\pm$29.39 & 2511.58$\pm$29.78\\
143018 & 3929.67$\pm$26.16 & 6605.39$\pm$33.14 & 2449.16$\pm$27.73\\
147165 & 739.23$\pm$15.18  & 1890.19$\pm$20.74 & 920.45$\pm$21.66\\
142669 & 798.11$\pm$18.98  & 1177.58$\pm$20.16 & 670.29$\pm$132.36\\
144470 & 562.69$\pm$14.21  & 1325.87$\pm$18.46 & 783.63$\pm$21.72\\
145502 & 215.81$\pm$12.88  & 537.18$\pm$15.70  & 230.42$\pm$22.33\\
145482 & 380.29$\pm$12.47  & 876.03$\pm$15.59  & 337.77$\pm$19.20\\
148605 & 306.32$\pm$13.00  & 687.86$\pm$15.92  & 316.78$\pm$20.62\\
141637 & 214.14$\pm$12.92  & 492.26$\pm$15.47  & 264.32$\pm$25.16\\
142114 & 338.71$\pm$12.42  & 882.97$\pm$15.85  & 501.40$\pm$20.30\\
147933 & 158.80$\pm$11.63  & 323.28$\pm$13.49  & 221.25$\pm$18.38\\
148479 & 99.14$\pm$11.41   & 320.55$\pm$13.66  & 172.12$\pm$20.34\\
\hline
148184 &  36.29$\pm$15.99  & 146.84$\pm$50.20  & 81.03$\pm$16.00\\
142096 &  44.85$\pm$23.61  & 146.73$\pm$53.54  & 89.88$\pm$17.75\\
142184 &  57.28$\pm$31.71  & 159.24$\pm$59.38  & 50.87$\pm$10.05\\
142990 &  20.92$\pm$9.80   & 80.90$\pm$29.08   & 50.46$\pm$9.96\\
142165 &  17.64$\pm$8.72   & 70.64$\pm$24.39   & 29.73$\pm$5.87\\
142378 &  10.75$\pm$4.89   & 44.85$\pm$14.54   & 29.88$\pm$5.90\\
142883 &  18.61$\pm$9.45   & 62.58$\pm$22.31   & 24.12$\pm$4.76\\
144334 &  0.06$\pm$0.03    & 2.42$\pm$0.87     & 21.60$\pm$4.26\\
142301 &  0.06$\pm$0.03    & 2.34$\pm$0.78     & 20.93$\pm$4.13\\
142250 &  4.56$\pm$2.13    & 23.99$\pm$8.62    & 16.86$\pm$3.33 \\
146001 &  1.03$\pm$0.51    & 8.85$\pm$3.08     & 12.94$\pm$2.56\\
145792 &  1.83$\pm$1.01    & 11.93$\pm$4.70    & 12.71$\pm$2.51\\
144661 &  0.35$\pm$0.16    & 4.40$\pm$1.58     & 12.64$\pm$2.50\\
145483 &  0.00$\pm$0.00    & 0.63$\pm$0.22     & 8.01$\pm$1.58\\
144844 &  0.04$\pm$0.02    & 1.87$\pm$0.69     & 10.71$\pm$2.11\\
142884 &  0.00$\pm$0.00    & 0.32$\pm$0.11     & 4.67$\pm$0.92 \\
146416 &  0.00$\pm$0.00    & 0.23$\pm$0.08     & 3.33$\pm$0.66\\
142315 &  0.03$\pm$0.02    & 1.14$\pm$0.41     & 3.45$\pm$0.68\\
145102 &  0.00$\pm$0.00    & 0.19$\pm$0.07     & 2.40$\pm$0.47 
\enddata
\end{deluxetable}

\clearpage

\begin{deluxetable}{lccccccc}
\tablecolumns{8}
\tablewidth{0pt}
\tablecaption{FM88 Curve Coefficients \label{FMCC}}
\tablehead{
\colhead{HD} & \colhead{Standard(HD)} & \colhead{$c_1$} & \colhead{$c_2$} & \colhead{$c_3$} 
& \colhead{$c_4$} & \colhead{$\gamma$} & \colhead{$\lambda_0^{-1}$}}
\startdata
149438 & ... & -0.07$\pm$0.02 & 0.70$\pm$0.01 & 3.23$\pm$0.05 & 0.41$\pm$0.01 & 0.99$\pm$0.01 & 4.596$\pm$0.011\\
143275 & 143018 & -1.29$\pm$0.22 & 0.86$\pm$0.04 & 2.59$\pm$0.35 & 0.22$\pm$0.03 & 0.77$\pm$0.06 & 4.555$\pm$0.014\\
144217 & 143018 & 1.07$\pm$0.15 & 0.38$\pm$0.03 & 2.73$\pm$0.19 & 0.32$\pm$0.02 & 0.66$\pm$0.03 & 4.502$\pm$0.007\\
143018 & ... & -0.07$\pm$0.02 & 0.70$\pm$0.01 & 3.23$\pm$0.05 & 0.41$\pm$0.01 & 0.99$\pm$0.01 & 4.596$\pm$0.011\\
147165 & 143018 & 1.46$\pm$0.08 & 0.29$\pm$0.01 & 2.77$\pm$0.14 & 0.15$\pm$0.01 & 0.83$\pm$0.02 & 4.572$\pm$0.006\\ 
142669 & ... & -0.07$\pm$0.02 & 0.70$\pm$0.01 & 3.23$\pm$0.05 & 0.41$\pm$0.01 & 0.99$\pm$0.01 & 4.596$\pm$0.011\\
144470 & 143018 & 0.10$\pm$0.15 & 0.65$\pm$0.03 & 3.43$\pm$0.27 & 0.12$\pm$0.03 & 0.81$\pm$0.03 & 4.536$\pm$0.008\\
145502 & 143018 & 1.28$\pm$0.12 & 0.50$\pm$0.02 & 3.58$\pm$0.26 & 0.34$\pm$0.03 & 0.90$\pm$0.03 & 4.526$\pm$0.008\\
145482 & ... & -0.07$\pm$0.02 & 0.70$\pm$0.01 & 3.23$\pm$0.05 & 0.41$\pm$0.01 & 0.99$\pm$0.01 & 4.596$\pm$0.011\\
148605 & ... & -0.07$\pm$0.02 & 0.70$\pm$0.01 & 3.23$\pm$0.05 & 0.41$\pm$0.01 & 0.99$\pm$0.01 & 4.596$\pm$0.011\\
141637 & 148605 & -1.29$\pm$0.17 & 0.73$\pm$0.03 & 2.11$\pm$0.27 & 0.17$\pm$0.08 & 0.77$\pm$0.05 & 4.554$\pm$0.013\\
142114 & 122980 & -0.54$\pm$0.20 & 0.64$\pm$0.04 & 4.05$\pm$0.43 & 0.30$\pm$0.10 & 0.89$\pm$0.05 & 4.590$\pm$0.012\\
147933 & 143018 & 0.89$\pm$0.06 & 0.40$\pm$0.01 & 2.83$\pm$0.01 & 0.17$\pm$0.02 & 0.83$\pm$0.02 & 4.529$\pm$0.004\\
148479 & 32630 & 2.40$\pm$0.20 & 0.22$\pm$0.01 & 2.88$\pm$0.37 & 0.68$\pm$0.09 & 0.83$\pm$0.06 & 4.567$\pm$0.013\\
148184 & 3360 & -0.35$\pm$0.06 & 0.51$\pm$0.01 & 3.32$\pm$0.11 & 0.13$\pm$0.03 & 0.84$\pm$0.01 & 4.538$\pm$0.003\\
142096 & 122980 & 1.22$\pm$0.17 & 0.53$\pm$0.03 & 4.11$\pm$0.51 & 0.34$\pm$0.07 & 1.08$\pm$0.06 & 4.586$\pm$0.014\\
142184 & 61831 & 1.42$\pm$0.17 & 0.50$\pm$0.03 & 1.38$\pm$0.19 & 0.30$\pm$0.09 & 0.63$\pm$0.05 & 4.581$\pm$0.012\\
142990 & ... & -0.07$\pm$0.02 & 0.70$\pm$0.01 & 3.23$\pm$0.05 & 0.41$\pm$0.01 & 0.99$\pm$0.01 & 4.596$\pm$0.011\\
142165 & 90994 & 0.99$\pm$0.25 & 0.43$\pm$0.05 & 2.45$\pm$0.45 & 0.03$\pm$0.11 & 0.82$\pm$0.08 & 4.583$\pm$0.019\\
142378 & 34759 & 0.15$\pm$0.22 & 0.33$\pm$0.04 & 4.53$\pm$0.55 & 0.06$\pm$0.09 & 0.99$\pm$0.06 & 4.532$\pm$0.013\\
142883 & 32630 & 0.65$\pm$0.14 & 0.58$\pm$0.03 & 2.14$\pm$0.21 & 0.28$\pm$0.07 & 0.73$\pm$0.04 & 4.560$\pm$0.009\\
144334 & ... & -0.07$\pm$0.02 & 0.70$\pm$0.01 & 3.23$\pm$0.05 & 0.41$\pm$0.01 & 0.99$\pm$0.01 & 4.596$\pm$0.011\\
142301 & 144334 & 0.35$\pm$0.25 & 0.56$\pm$0.05 & 2.80$\pm$0.51 & 0.29$\pm$0.12 & 0.85$\pm$0.08 & 4.601$\pm$0.019\\
142250 & ... & -0.07$\pm$0.02 & 0.70$\pm$0.01 & 3.23$\pm$0.05 & 0.41$\pm$0.01 & 0.99$\pm$0.01 & 4.596$\pm$0.011\\
146001 & 29335 & -0.02$\pm$0.21 & 0.56$\pm$0.04 & 4.11$\pm$0.59 & 0.07$\pm$0.09 & 1.04$\pm$0.07 & 4.534$\pm$0.016\\
145792 & 34759 & 1.37$\pm$0.19 & 0.33$\pm$0.03 & 3.15$\pm$0.40 & 0.56$\pm$0.08 & 0.92$\pm$0.06 & 4.507$\pm$0.013\\
144661 & ... & -0.07$\pm$0.02 & 0.70$\pm$0.01 & 3.23$\pm$0.05 & 0.41$\pm$0.01 & 0.99$\pm$0.01 & 4.596$\pm$0.011\\
145483 & ... & -0.07$\pm$0.02 & 0.70$\pm$0.01 & 3.23$\pm$0.05 & 0.41$\pm$0.01 & 0.99$\pm$0.01 & 4.596$\pm$0.011\\
144844 & 144334\tablenotemark{a} & -0.07$\pm$0.02 & 0.70$\pm$0.01 & 3.23$\pm$0.05 & 0.41$\pm$0.01 & 0.99$\pm$0.01 & 4.596$\pm$0.011\\
142884 & 142184\tablenotemark{a} &  1.42$\pm$0.17 & 0.50$\pm$0.03 & 1.38$\pm$0.19 & 0.30$\pm$0.09 & 0.63$\pm$0.05 & 4.581$\pm$0.012\\
146416 & ... & -0.07$\pm$0.02 & 0.70$\pm$0.01 & 3.23$\pm$0.05 & 0.41$\pm$0.01 & 0.99$\pm$0.01 & 4.596$\pm$0.011\\
142315 & 222173 & 1.69$\pm$0.24 & 0.18$\pm$0.05 & 3.08$\pm$0.46 & 0.27$\pm$0.11 & 0.85$\pm$0.07 & 4.604$\pm$0.016\\
145102 & 196867 & -1.16$\pm$0.21 & 0.70$\pm$0.04 & 3.08$\pm$0.44 & 0.17$\pm$0.10 & 0.91$\pm$0.07 & 4.509$\pm$0.015
\enddata
\tablenotetext{a}{Derived from nearest neighbor}
\end{deluxetable}

\clearpage

\begin{deluxetable}{lcccc}
\tablecolumns{5}
\tablewidth{0pt}
\tablecaption{$R_V$ Values and Stellar Optical Depths\label{tautab}}
\tablehead{\colhead{HD} & \colhead{$R_V$} & \colhead{$\tau_{eff}$[973\AA]} &
\colhead{$\tau_{eff}$[1106\AA]} & \colhead{$\tau_{eff}$[1261\AA]}}
\startdata
149438 & 3.10$\pm$0.10 & 0.515$\pm$0.332  & 0.406$\pm$0.263 & 0.296$\pm$0.191\\
143275 & 2.96$\pm$0.48 & 1.713$\pm$0.293  & 1.340$\pm$0.254 & 0.978$\pm$0.188\\
144217 & 2.59$\pm$0.26 & 2.224$\pm$0.259  & 1.640$\pm$0.217 & 1.161$\pm$0.157\\
143018 & 3.10$\pm$0.10 & 1.152$\pm$0.339  & 0.847$\pm$0.286 & 0.567$\pm$0.192\\
147165 & 3.53$\pm$0.19 & 3.463$\pm$0.216  & 2.896$\pm$0.197 & 2.393$\pm$0.166\\
142669 & 3.10$\pm$0.10 & 0.455$\pm$0.342  & 0.320$\pm$0.320 & 0.192$\pm$0.192\\
144470 & 3.23$\pm$0.24 & 2.440$\pm$0.254  & 1.942$\pm$0.230 & 1.531$\pm$0.183\\
145502 & 3.28$\pm$0.23 & 3.636$\pm$0.306  & 2.882$\pm$0.261 & 2.109$\pm$0.194\\
145482 & 3.10$\pm$0.10 & 0.827$\pm$0.339  & 0.607$\pm$0.303 & 0.384$\pm$0.192\\
148605 & 3.10$\pm$0.10 & 1.114$\pm$0.325  & 0.858$\pm$0.273 & 0.590$\pm$0.188\\
141637 & 2.24$\pm$0.34 & 1.755$\pm$0.310  & 1.404$\pm$0.207 & 1.017$\pm$0.150\\
142114 & 2.51$\pm$0.29 & 1.962$\pm$0.413  & 1.429$\pm$0.238 & 1.012$\pm$0.165\\
147933 & 4.07$\pm$0.15 & 5.404$\pm$0.249  & 4.301$\pm$0.220 & 3.485$\pm$0.182\\
148479 & 3.10$\pm$0.10 & 2.523$\pm$0.433  & 1.881$\pm$0.281 & 1.169$\pm$0.172\\
148184 & 4.23$\pm$0.14 & 5.566$\pm$0.251  & 4.532$\pm$0.215 & 3.703$\pm$0.178\\
142096 & 3.34$\pm$0.25 & 3.041$\pm$0.375  & 2.366$\pm$0.269 & 1.758$\pm$0.200\\
142184 & 3.91$\pm$0.38 & 2.388$\pm$0.408  & 1.878$\pm$0.278 & 1.410$\pm$0.208\\
142990 & 3.10$\pm$0.10 & 1.536$\pm$0.328  & 1.152$\pm$0.281 & 0.781$\pm$0.191\\
142165 & 4.11$\pm$0.51 & 1.305$\pm$0.375  & 1.078$\pm$0.228 & 0.892$\pm$0.183\\
142378 & 3.98$\pm$0.40 & 1.377$\pm$0.315  & 1.110$\pm$0.192 & 0.899$\pm$0.152\\
142883 & 3.30$\pm$0.25 & 2.861$\pm$0.358  & 2.246$\pm$0.256 & 1.675$\pm$0.191\\
144334 & 3.10$\pm$0.10 & 1.361$\pm$0.321  & 0.938$\pm$0.254 & 0.701$\pm$0.191\\
142301 & 1.80$\pm$0.31 & 1.417$\pm$0.409  & 0.968$\pm$0.209 & 0.728$\pm$0.151\\
142250 & 3.10$\pm$0.10 & 0.896$\pm$0.328  & 0.654$\pm$0.287 & 0.433$\pm$0.190\\
146001 & 4.04$\pm$0.38 & 1.804$\pm$0.362  & 1.467$\pm$0.226 & 1.216$\pm$0.184\\
145792 & 4.87$\pm$0.39 & 2.944$\pm$0.417  & 2.221$\pm$0.302 & 1.569$\pm$0.214\\
144661 & 3.10$\pm$0.10 & 1.612$\pm$0.321  & 1.177$\pm$0.267 & 0.840$\pm$0.191\\
145483 & 3.10$\pm$0.10 & 1.054$\pm$0.332  & 0.655$\pm$0.244 & 0.504$\pm$0.188\\
144844 & 3.31$\pm$0.43 & 1.911$\pm$0.324  & 1.353$\pm$0.260 & 1.024$\pm$0.200\\
142884 & 2.11$\pm$0.24 & 2.121$\pm$0.374  & 1.476$\pm$0.212 & 1.173$\pm$0.167\\
146416 & 3.10$\pm$0.10 & 1.309$\pm$0.331  & 0.831$\pm$0.244 & 0.642$\pm$0.188\\
142315 & 4.79$\pm$0.57 & 1.522$\pm$0.392  & 1.108$\pm$0.229 & 0.911$\pm$0.185\\
145102 & 4.29$\pm$0.42 & 1.964$\pm$0.416  & 1.406$\pm$0.230 & 1.183$\pm$0.191
\enddata
\end{deluxetable}

\clearpage

\begin{deluxetable}{lccc}
\tablecolumns{4}
\tablewidth{0pt} 
\tablecaption{Dereddened Luminosities\label{deredlum}} 
\tablehead{ \colhead{HD} & \colhead{$L_0$[973\AA]} & \colhead{$L_0$[1106\AA]} & \colhead{$L_0$[1261\AA]}\\
\colhead{} & \colhead{(counts $\rm s^{-1}$ $\rm bandpass^{-1}$)} 
& \colhead{(counts $\rm s^{-1}$ $\rm bandpass^{-1}$)} 
& \colhead{(counts $\rm s^{-1}$ $\rm bandpass^{-1}$)}}
\startdata
149438 &  10245.66$\pm$3441.07 &  10425.00$\pm$2562.57 &  4868.16$\pm$888.83\\
143275 &  25820.66$\pm$7523.29 &  28877.76$\pm$6605.25 &  8562.69$\pm$1553.04\\
144217 &  22075.63$\pm$5718.65 &  24581.37$\pm$4746.53 &  6592.99$\pm$987.86\\
143018 &  12639.08$\pm$4245.47 &  15448.11$\pm$3798.14 &  3865.27$\pm$706.00\\
147165 &  24788.61$\pm$5397.26 &  34514.40$\pm$6397.50 & 11254.84$\pm$1837.60\\
142669 &   1154.06$\pm$387.59  &   1636.54$\pm$402.26  &   812.09$\pm$148.26\\
144470 &   6821.63$\pm$1717.75 &   9307.61$\pm$1946.01 &  3750.34$\pm$667.08\\
145502 &   9356.78$\pm$2939.32 &   9679.58$\pm$2302.78 &  2161.15$\pm$408.95\\
145482 &    925.42$\pm$310.82  &   1591.84$\pm$391.31  &   552.94$\pm$100.97\\
148605 &   1020.50$\pm$342.79  &   1591.08$\pm$391.21  &   732.11$\pm$133.73\\
141637 &   1369.92$\pm$455.73  &   2025.88$\pm$388.45  &   496.32$\pm$72.53\\
142114 &   2621.35$\pm$1118.08 &   3696.17$\pm$803.85  &  1380.95$\pm$214.42\\
147933 &  43881.32$\pm$10957.16&  24816.03$\pm$5148.69 & 12942.74$\pm$2326.49\\
148479 &   1613.21$\pm$738.08  &   2050.59$\pm$516.89  &   794.95$\pm$134.24\\
\hline
148184 &  10002.99$\pm$3585.40 &  14499.43$\pm$3999.70 &  3300.90$\pm$873.62\\
142096 &    959.72$\pm$337.73  &   1608.49$\pm$429.98  &   522.12$\pm$145.48\\
142184 &    632.95$\pm$224.99  &   1060.83$\pm$288.53  &   208.64$\pm$59.45\\
142990 &     98.07$\pm$33.11   &    259.12$\pm$67.95   &   110.27$\pm$30.11\\
142165 &     65.45$\pm$22.60   &    208.89$\pm$57.02   &    72.51$\pm$19.49\\
142378 &     42.96$\pm$14.80   &    137.09$\pm$37.27   &    73.45$\pm$18.26\\
142883 &    331.62$\pm$115.84  &    606.69$\pm$162.06  &   128.96$\pm$35.15\\
144334 &      0.24$\pm$0.08    &     6.23$\pm$1.63     &    43.57$\pm$11.90\\
142301 &      0.24$\pm$0.08    &     6.21$\pm$1.61     &    43.35$\pm$10.71\\
142250 &     11.22$\pm$3.78    &    46.41$\pm$12.17    &    26.00$\pm$7.10\\
146001 &      6.33$\pm$2.20    &    38.75$\pm$10.56    &    43.65$\pm$11.75\\
145792 &     35.59$\pm$12.48   &   113.58$\pm$31.76    &    61.12$\pm$17.70\\
144661 &      1.75$\pm$0.60    &    14.43$\pm$3.78     &    29.31$\pm$8.00\\
145483 &      0.01$\pm$0.00    &     1.22$\pm$0.31     &    13.26$\pm$3.62\\
144844 &      0.28$\pm$0.10    &     7.30$\pm$1.95     &    29.85$\pm$8.35\\
142884 &      0.02$\pm$0.01    &     1.39$\pm$0.36     &    15.10$\pm$3.89\\
146416 &      0.00$\pm$0.00    &     0.54$\pm$0.14     &     6.32$\pm$1.73\\
142315 &      0.13$\pm$0.05    &     3.47$\pm$0.97     &     8.58$\pm$2.31\\
145102 &      0.01$\pm$0.00    &     0.78$\pm$0.21     &     7.83$\pm$2.14
\enddata
\end{deluxetable}

\clearpage
{\plotone{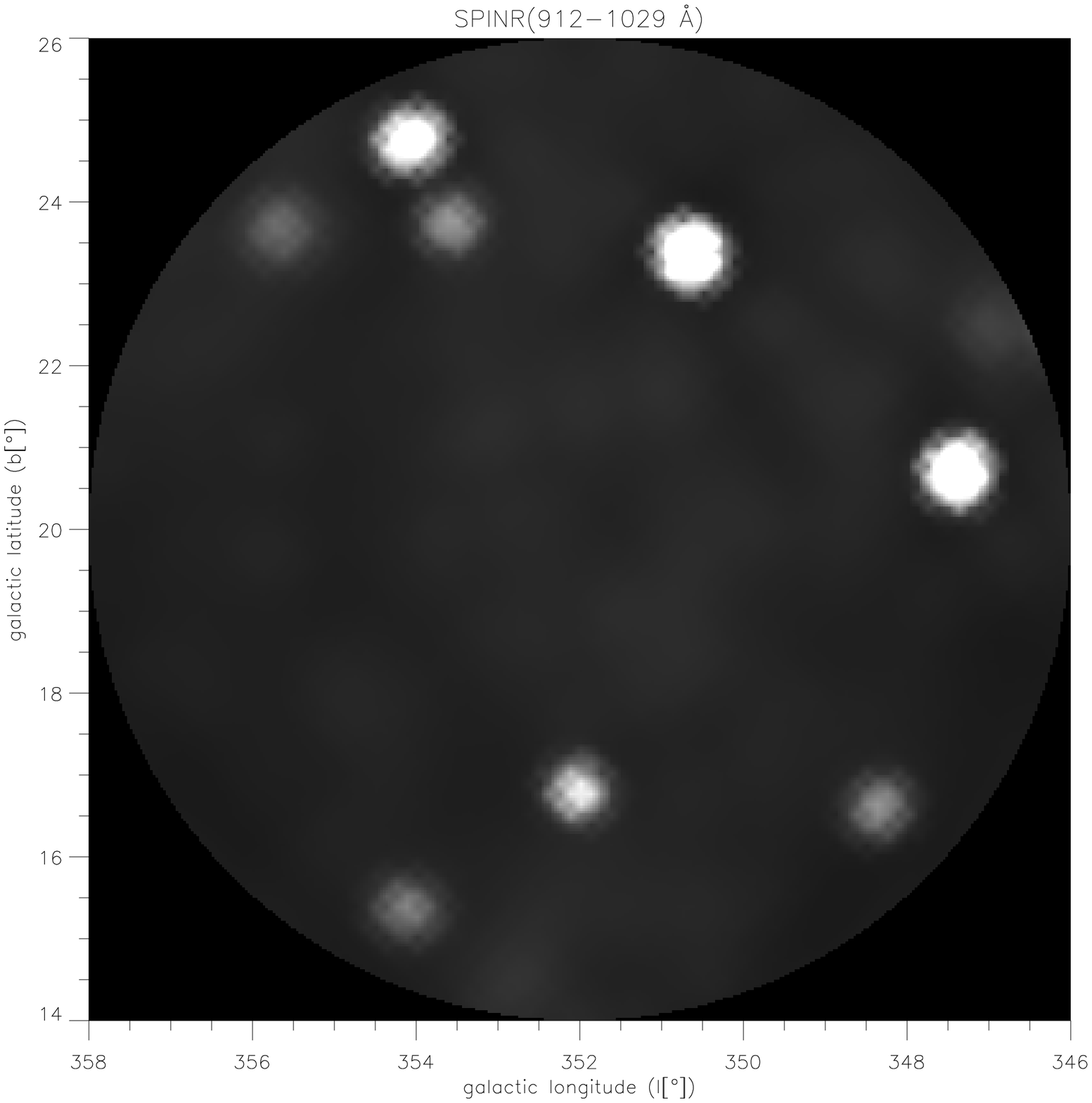}}
\clearpage
{\plotone{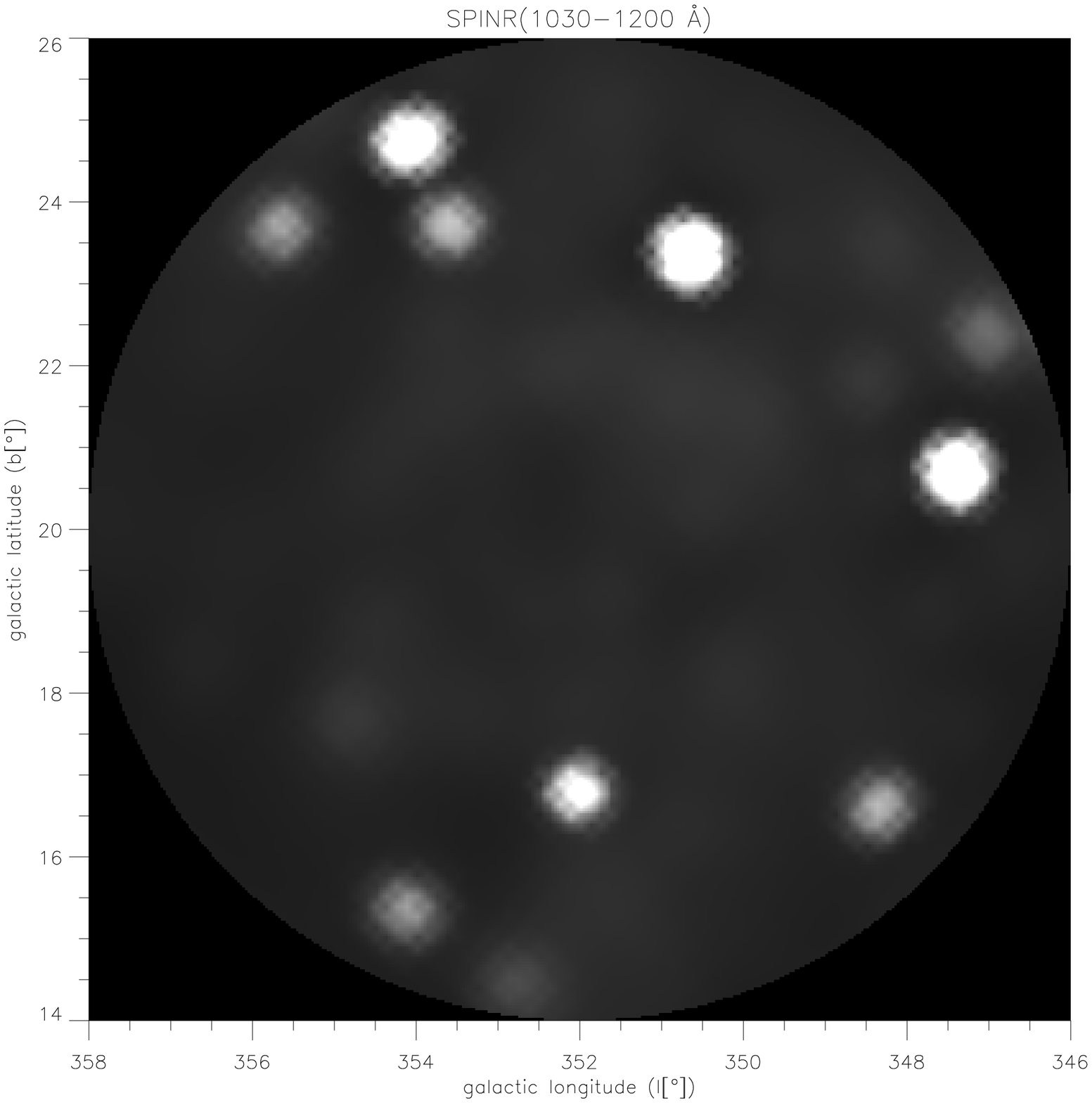}}
\clearpage
{\plotone{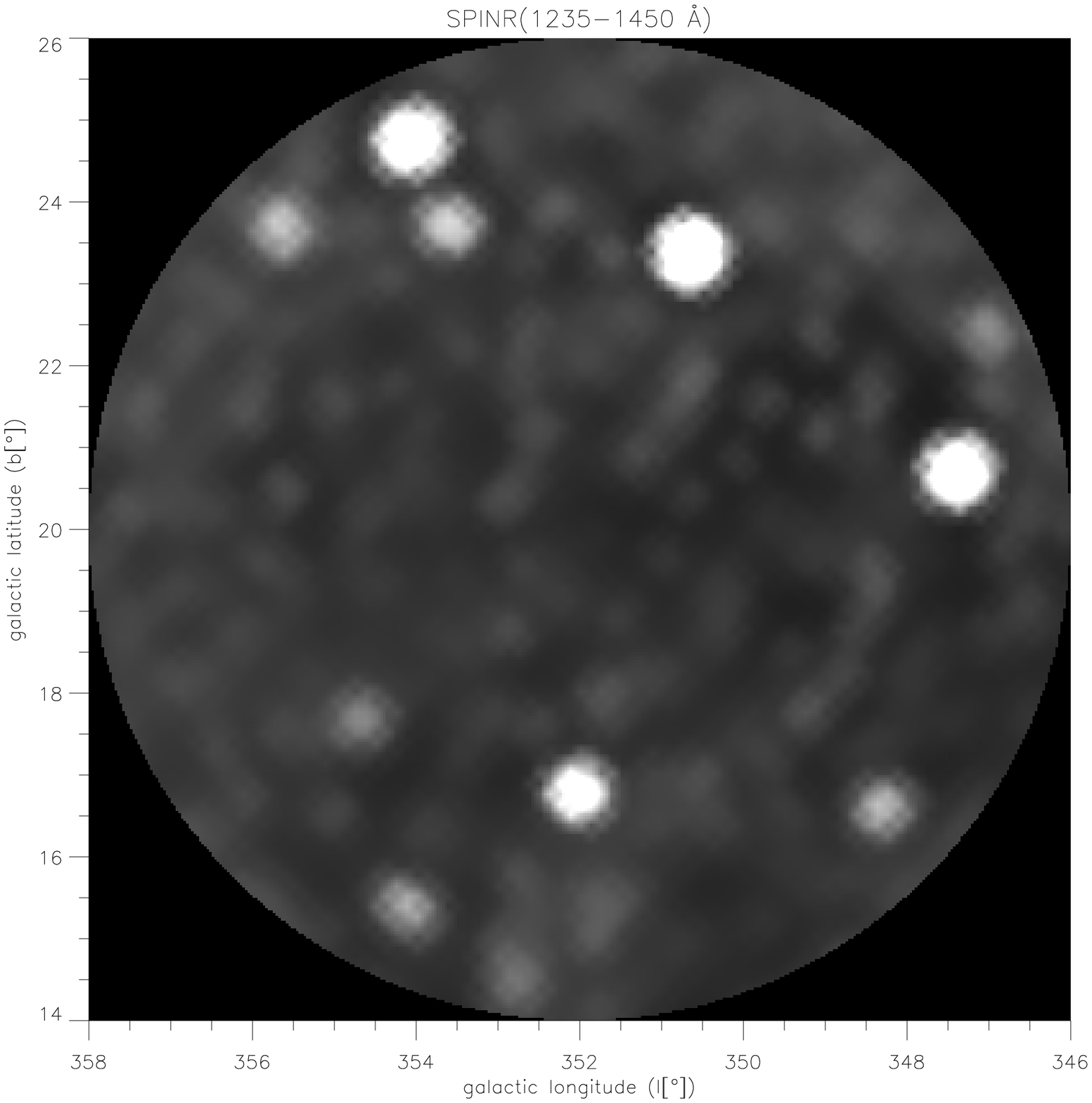}}
\clearpage

\clearpage

\clearpage

\clearpage

\clearpage
{\plotone{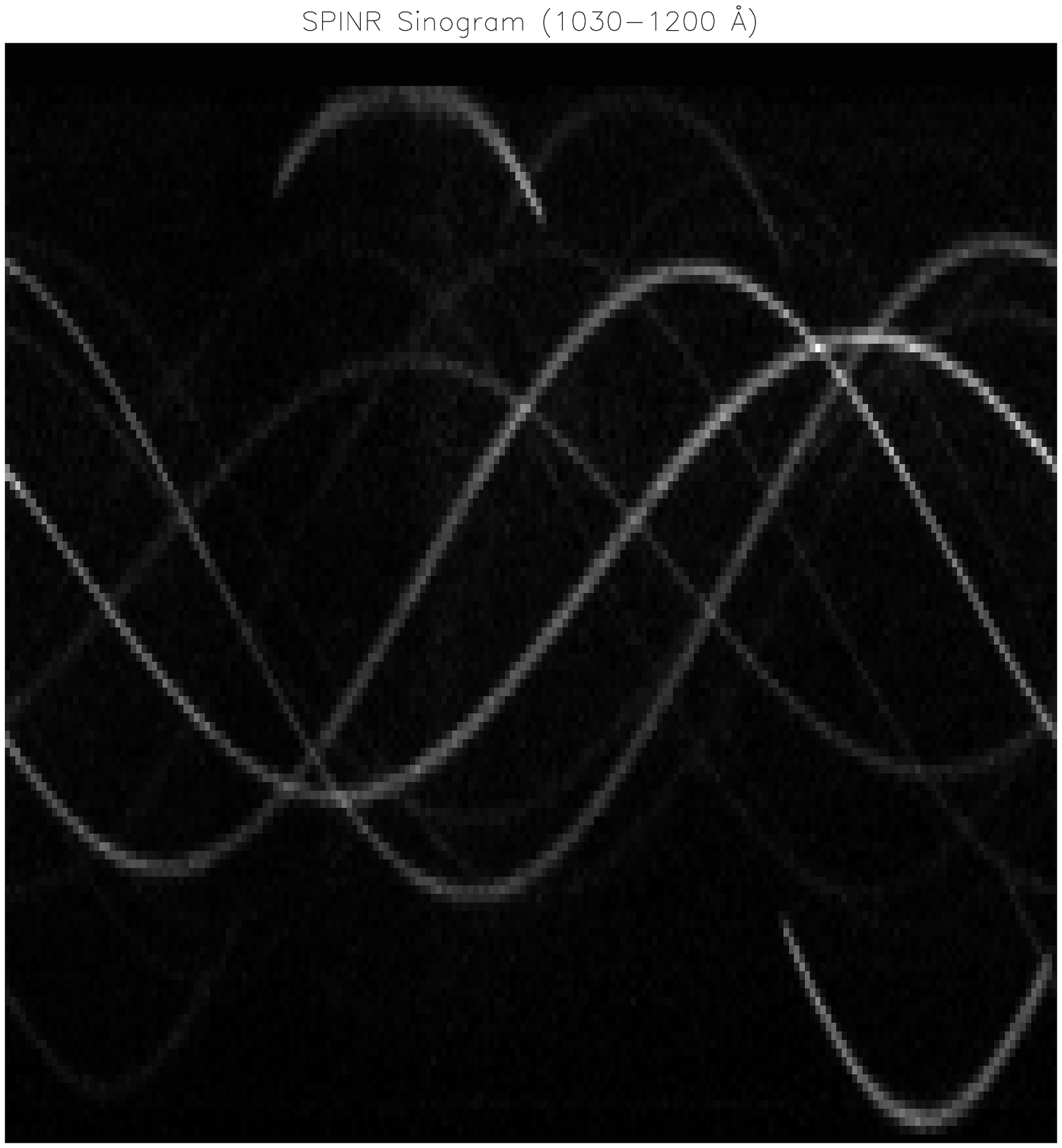}}
\clearpage
{\plotone{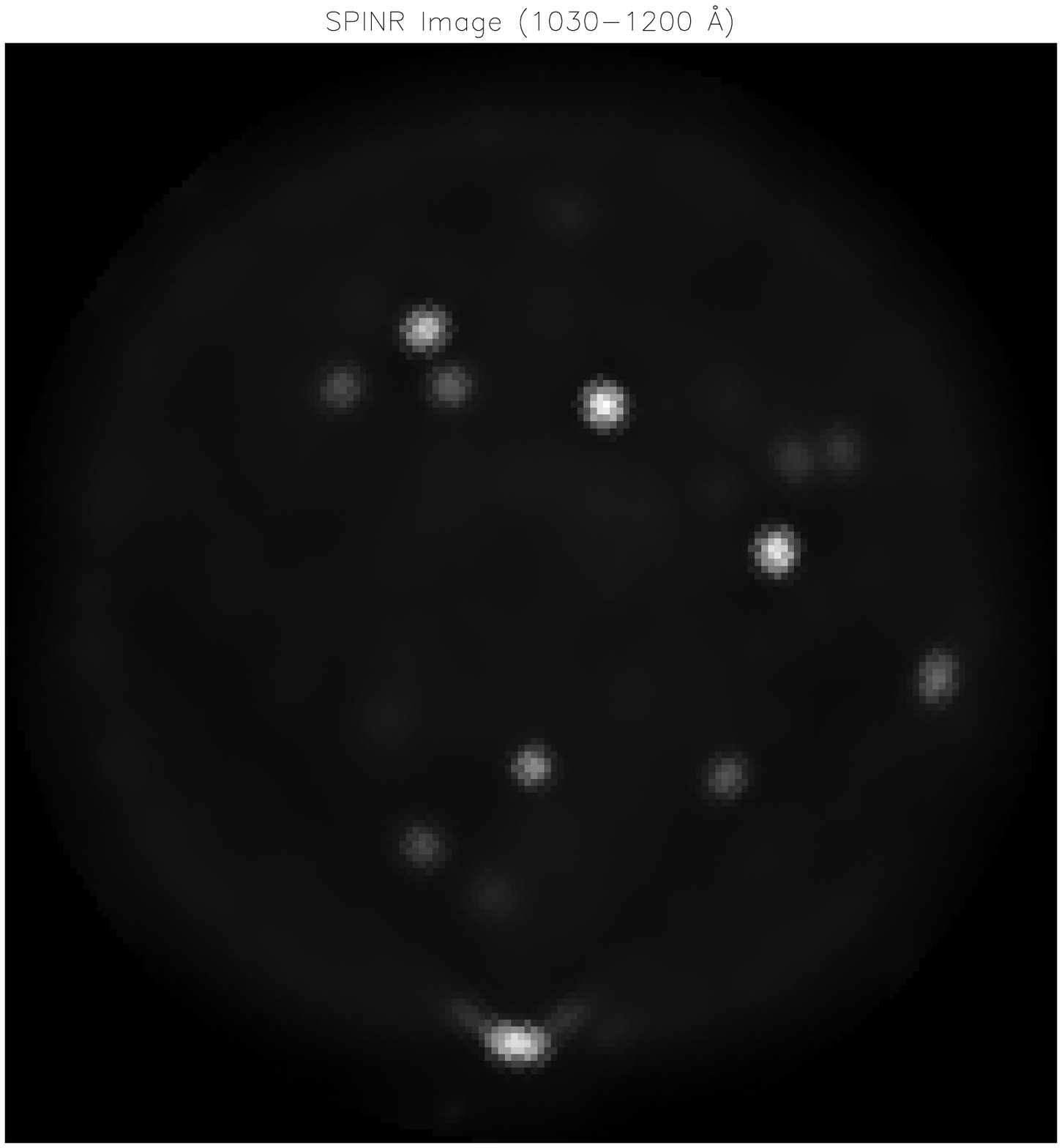}}
\clearpage
{\plotone{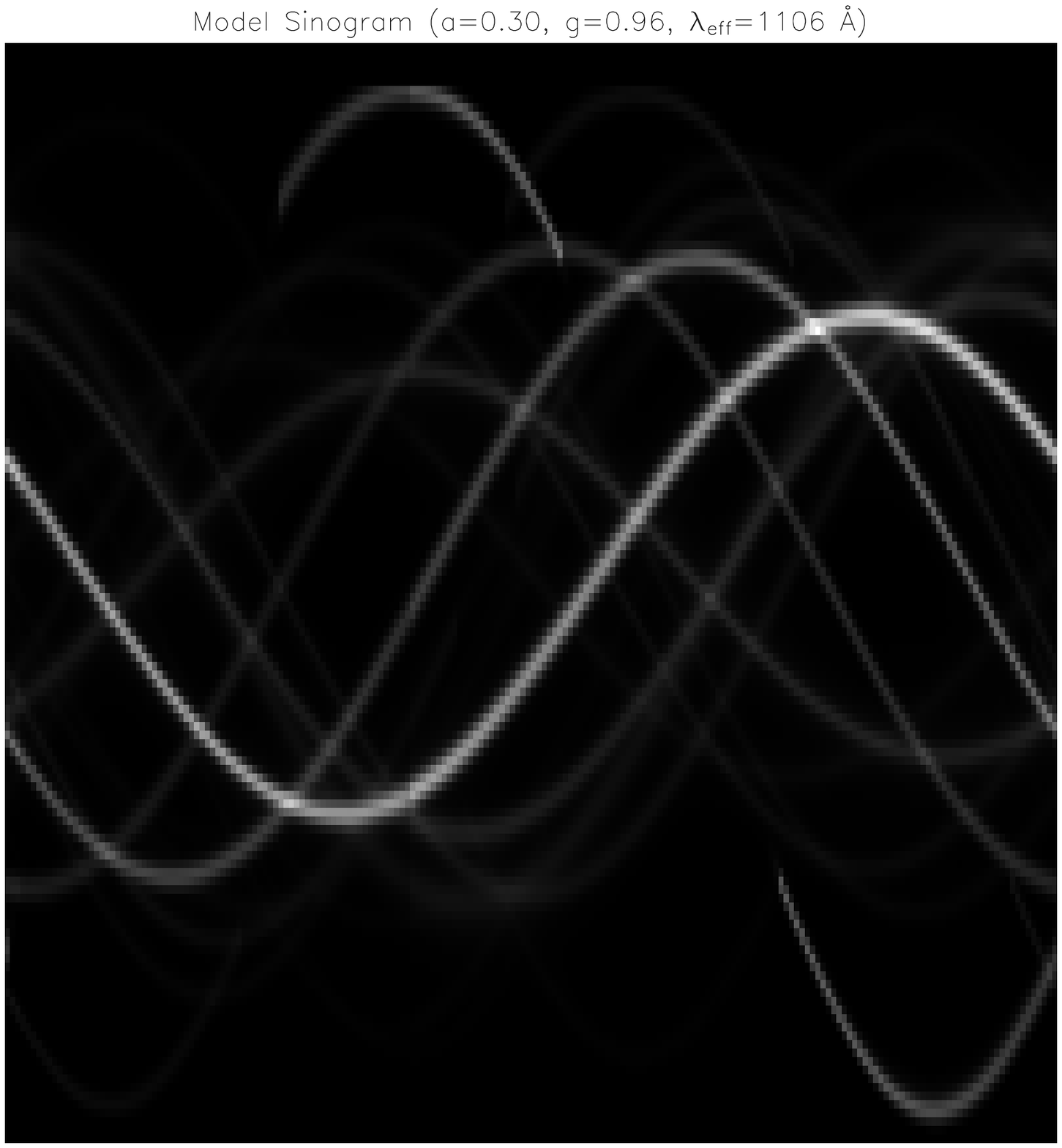}}
\clearpage
{\plotone{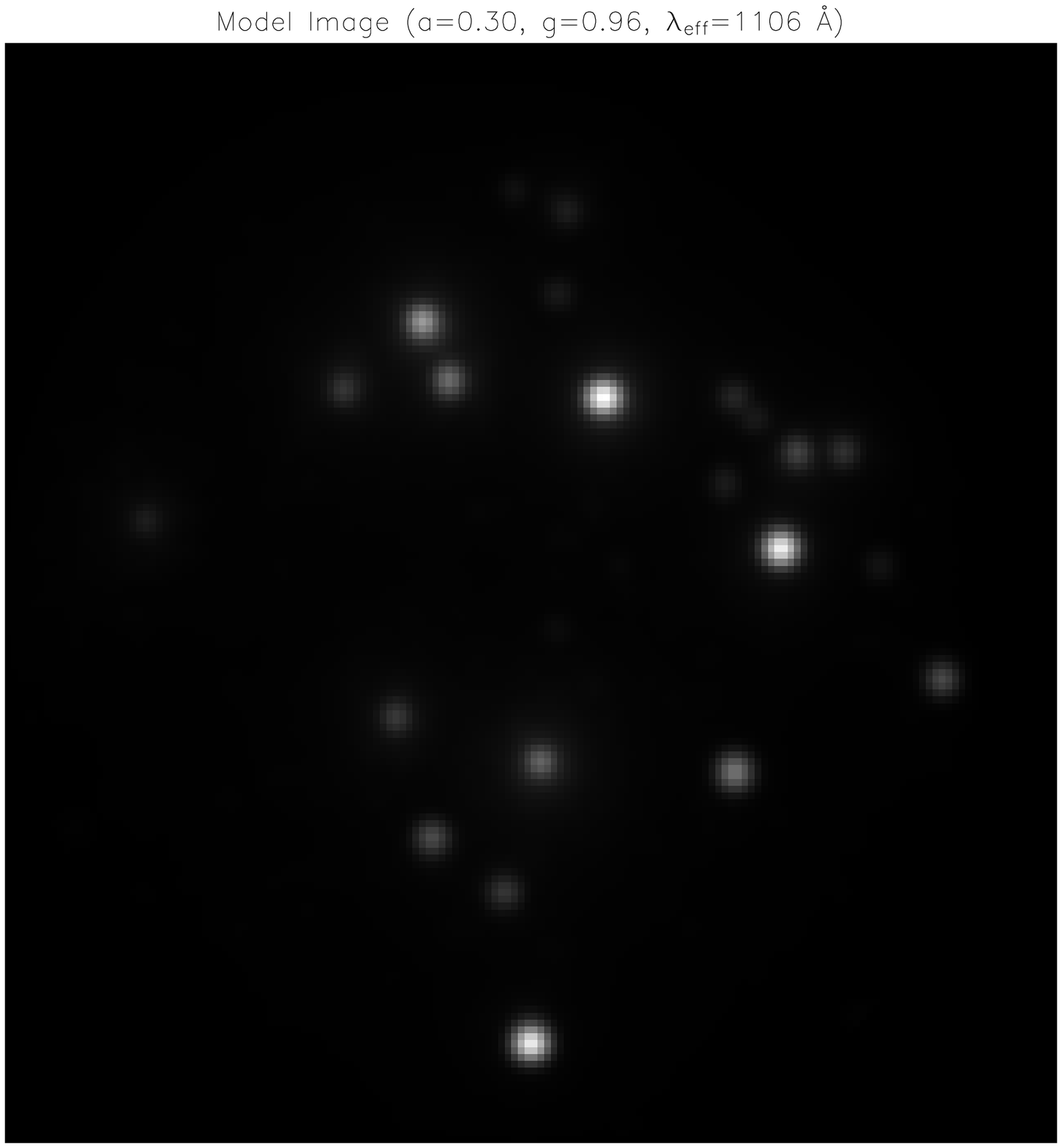}}
\clearpage
{\plotone{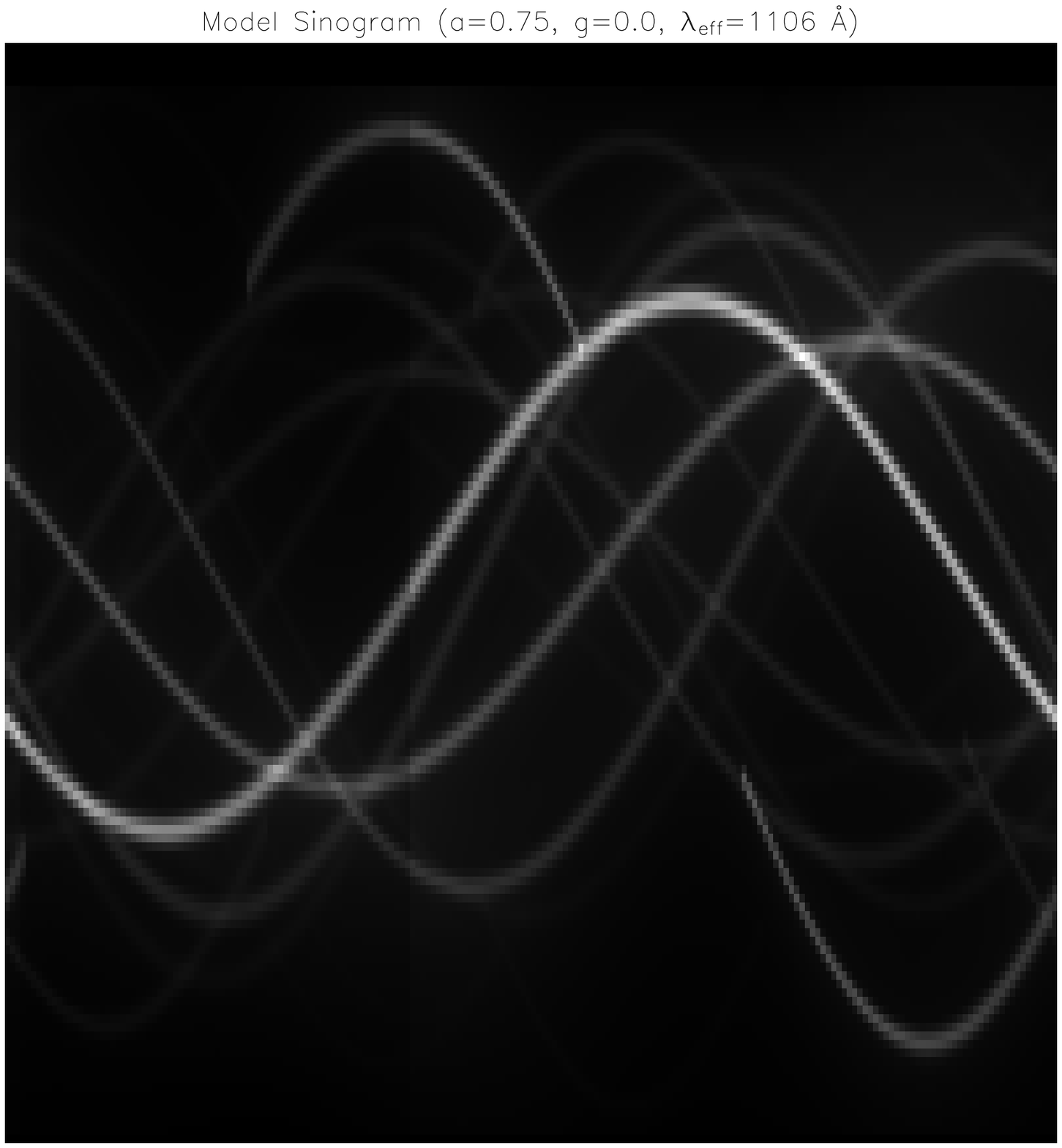}}
\clearpage

\clearpage

\end{document}